\documentclass[ba]{imsart}
\pubyear{2026}
\volume{TBA}
\issue{TBA}
\firstpage{1}
\lastpage{1}

\usepackage{tikz}
\usepackage{amsmath}
\usepackage{amssymb}
\usepackage{amsthm}
\usepackage{mathtools}

\DeclareMathOperator*{\argmin}{argmin}

\usepackage{natbib}
\usepackage[colorlinks,citecolor=blue,urlcolor=blue,filecolor=blue,backref=page]{hyperref}
\usepackage{graphicx}
\newcommand{\capitalhyphen}{\raisebox{0.24ex}{\resizebox{0.4em}{\height}{-}}\kern-0.07em}

\usepackage{fancyhdr}

\startlocaldefs
\endlocaldefs

\startlocaldefs
\theoremstyle{plain}

\theoremstyle{definition}

\theoremstyle{remark}

\endlocaldefs

\usepackage{ragged2e}
\usepackage{rotating}

\usepackage{algorithm}%
\usepackage[algo2e]{algorithm2e}%

\usepackage{listings}
\usepackage{color}
\definecolor{dkgreen}{rgb}{0,0.6,0}
\definecolor{gray}{rgb}{0.5,0.5,0.5}
\definecolor{mauve}{rgb}{0.58,0,0.82}
\usepackage{soul}

\usepackage{booktabs}
\usepackage{caption}
\usepackage{subcaption}

\begin{document}


\begin{frontmatter}
\title{Distributed and recursive Bayesian inference for Big Data and complex spatio-temporal models}

\runtitle{}

\begin{aug}
\author{\fnms{Mario} \snm{Figueira}\thanksref{addr1,t1}\ead[label=e1]{Mario.Figueira-Pereira@vogelwarte.ch}},
\author{\fnms{David} \snm{Conesa}\thanksref{addr2}\ead[label=e2]{}},
\author{\fnms{Antonio} \snm{López-Quílez}\thanksref{addr2}\ead[label=e3]{}} \and
\author{\fnms{H\r{a}vard} \snm{Rue}\thanksref{addr2}\ead[label=e4]{}}

\runauthor{}

\address[1]{Population Biology Research Unit, Swiss Ornithological Institute. Sempach (Luzern), Switzerland.}

\address[addr2]{Department of Statistics and Operations Research. Faculty of Mathematics C/ Dr. Moliner, 50 46100 Burjassot (València).}

\address[addr2]{CEMSE Division, King Abdullah University of Science and Technology, Kingdom of Saudi Arabia.}

\thankstext{t1}{\printead{e1}}

\end{aug}

\begin{abstract}

The rapid growth of massive and complex datasets in fields such as econometrics, environmental sciences, risk management, and public policy has reshaped statistical modeling while introducing significant computational and methodological challenges. These challenges arise not only from data scale and model complexity, but also from the sequential or streaming nature of modern applications and from data-privacy constraints that prevent sharing raw data and thus limit joint analysis. To address these challenges, we introduce a novel and comprehensive Bayesian framework for distributed and recursive inference, grounded in the Integrated Nested Laplace Approximations (INLA) methodology and implemented using the \texttt{R-INLA} software. Our contributions include the partitioning both data and structured model components, reducing computational complexity while preserving accuracy relative to centralized full-data inference. We demonstrate the effectiveness of the proposed framework through case studies that highlight its applicability in large-scale, streaming, and privacy-sensitive settings. By integrating distributed, federated, and recursive paradigms, this work offers scalable, adaptive, and generalizable tools for modern Bayesian inference.

\end{abstract}


\end{frontmatter}

\section{Introduction}\label{Introduction}

The increasing availability of large and complex datasets in fields such as econometrics, environmental sciences, ecology, risk management, and public policy has transformed the way statistical models are built and applied \citep{Varian_2014_BigDataEconometrics, Farley_2018_BigDataEcology, Sebestyen_2021_BigDataEnvironment}. The ability to measure, store, and access these extensive databases provides comprehensive insights into various phenomena while also posing significant challenges for statistical and computational theory and methodology, which motivates the development of distributed methods capable of harnessing the collective computational power of multiple machines or servers \citep{Gao_2022_ReviewDistributed, Zhou_2024_BigDataDistributed}. These challenges arise not only from the sheer volume of data but also from the complexity of models, such as spatial and spatio-temporal models. Modern data are often not only massive in size but also arrive sequentially or in streams, requiring models to be updated regularly as new information becomes available. Re-analyzing the entire dataset each time new data are collected quickly becomes computationally infeasible, particularly for high-dimensional or spatio-temporal models. This challenge motivates the development of recursive inference methods, which enable the posterior distribution of model parameters to be updated incrementally, incorporating new data without revisiting the full historical dataset. Such approaches are crucial not only for real-time decision-making, continuous monitoring and learning, or long-term data accumulation \citep{Hooten_2021_RecursiveInference, Kessler_2023_SeqBConLearning}, but also for meta-analysis \citep{Lunn_2013_RecursiveMetaAnalysis}, where evidence from new studies is published over time and must be integrated into an existing synthesis without restarting the full inferential process. In all these contexts, scalability and adaptability are as important as statistical rigor.

The challenges previously addressed have given rise to the development of a broad spectrum of distributed and recursive inference methodologies \citep{Huang_2005_SamplingLargeData, Vehtari_2020_ExpectationWayLife, Hooten_2021_RecursiveInference, Gao_2022_ReviewDistributed, Zhou_2024_BigDataDistributed}. Distributed approaches, typically based on divide-and-conquer strategies \citep{Huang_2005_SamplingLargeData, Vehtari_2020_ExpectationWayLife, Orozco_2023_BigDM}, partition the likelihood into subsets of data that can be analyzed in parallel and subsequently recombined using methods such as consensus Monte Carlo \citep{Scott_2016_ConsensusMonteCarlo, Rendell_2021_GlobalConsensus, Figueira_2025_SCB}, Weierstrass samplers \citep{Wang_2013_ParallelWeierstrass}, or Wasserstein barycenter aggregation of subposteriors \citep{Srivastava_2018_ScalableBayesWass}, among others. Recursive inference, in contrast, emphasizes sequential updating of posterior distributions as new data become available \citep{Figueira_2025_SCB, Figueira_2025_RecursiveInference}, avoiding the need to reprocess the entire dataset. Methods such as streaming variational Bayes \citep{Broderick_2013_StreamingVB}, stochastic variational inference \citep{Hoffman_2013_StochasticVI}, and particle-based sequential updates \citep{Carvalho_2010_ParticleLearning} enable efficient learning in settings where data arrive continuously or in large batches. These algorithmic frameworks are especially valuable in real-time applications, including environmental monitoring, financial risk assessment, and epidemiological surveillance. Moreover, beyond computational efficiency, distributed and recursive methods address increasingly important constraints on data accessibility. In privacy-preserving environments ---such as health care, business analytics, or policy evaluation--- data cannot always be centralized \citep{Sarwate_2013_DifferentialPrivacy, Smith_2017_InteractionDistributedPrivacy, Altman_2018_BigDataPrivacy}. Federated inference frameworks extend distributed computation by enabling collaborative learning across decentralized data sources without direct exchange of raw information, instead relying on communication of model summaries, gradients, or approximate posterior components \citep{Mcmahan_2017_FederatedLearning, Smith_2017_InteractionDistributedPrivacy, Yadav_2022_FederatedLearning}. This integration of distributed, recursive, and federated paradigms offers not only scalable solutions to big-data problems but also principled mechanisms for updating posterior information adaptively and responsibly in dynamic, heterogeneous, and privacy-sensitive environments.

In this paper, we present a comprehensive framework for performing distributed and recursive inference within the Bayesian paradigm. Our proposals build upon the \textit{integrated nested Laplace approximation} \citep[INLA,][]{Rue_2009_INLA, Niekerk_2023_INLAv2} methodology, taking advantage of the high computational performance provided by the \texttt{R-INLA} software \citep{Martins_2013_INLAsoftware}. This efficiency arises from the integration of sparse matrix algebra techniques ---particularly effective for Gaussian Markov Random Fields (GMRFs)--- and advanced numerical methods for deterministic approximation of posterior distributions. 

In particular, in the context of distributed inference, we provide a detailed explanation of both the conceptual foundations and the practical strategies needed to implement our proposal. These methods are not only effective for large-scale datasets and complex models, such as spatio-temporal structures, but also crucial in privacy-sensitive settings, where raw data cannot be shared. In such cases, joint inference is achieved through the exchange of summaries or aggregated information, preserving confidentiality while enabling rigorous analysis. Two complementary paradigms illustrate this perspective: a top-down approach, motivated by the need to scale inference to massive datasets through data and likelihood partitioning; and a bottom-up approach, where data are naturally split across sources, as in privacy-preserving computation or meta-analysis. Together, these views underscore the versatility of distributed inference as both a computational solution and a methodological framework in Bayesian analysis.

Within the framework of recursive inference, we also provide a detailed discussion of the conceptual foundations and limitations of our proposal, as well as practical implementation strategies. Recursive methods are motivated by scenarios where data arrive sequentially or are too large to be processed in a single step, requiring inference to be updated incrementally as new information becomes available. This setting is especially relevant in online learning, real-time monitoring, and streaming data applications, where recomputing the posterior from scratch is computationally prohibitive. By reusing information from previous inference steps, recursive approaches balance accuracy and efficiency, enabling near real-time updates while keeping computational costs manageable. However, these methods face challenges such as error accumulation and approximation quality, which call for careful algorithmic design. Overall, recursive inference offers a principled way to adapt Bayesian analysis to dynamic and evolving data environments.

In distributed and recursive approaches, a central challenge is how to partition the data and the model's structured components. Effective partitioning is essential for any distributed or recursive method, as it must balance computational efficiency with inferential accuracy. To address this, we leverage algorithms for the automatic partitioning of both data and structured effects, reducing computational complexity while limiting divergence from full-data inference. We then illustrate the performance of the proposed strategies through examples demonstrating their integration within distributed and recursive frameworks.

In summary, Section 2 provides a brief overview of the INLA methodology. Section 3 presents our proposed distributed inference approach within the INLA framework, offering a detailed discussion of various strategies and their limitations. Section 4 describes the recursive inference methods we propose, examining implementation strategies and addressing key methodological aspects. Section 5 introduces the partitioning of the latent field, its integration within both framework and its implications. Section 6 illustrates the application of these methods using one simulated case study and two real-world examples. In the conclusions we synthesize our proposals, highlighting their scope, applicability, and potential impact. Finally, in the \emph{Supplementary Material} we detailed several algorithms for the automatic partitioning of data and models, and we discuss the implementation of distributed and recursive inference within the \texttt{R-INLA} software, concisely connecting theory and practice.

\section{Integrated Nested Laplace Approximations}

INLA is a deterministic approximate Bayesian inference approach grounded in the properties of the GMRF and its computational efficiency. This methodology was developed by \cite{Rue_2009_INLA}, implemented in \texttt R through the \texttt {R-INLA package} \citep{Martins_2013_INLAsoftware}, and later extended to more flexible modeling structures \citep{Lindgren_2015_INLASpatialReview, Bakka_2018_INLAreview, Virgilio_2020_INLAbook}. This approach allows Bayesian inference to be performed on a wide range of structured additive models, known as Latent Gaussian Models (LGMs). While INLA primarily focuses on computing marginal posterior distributions, it effectively addresses the challenges of determining the joint posterior distribution of the latent field $\mathbf{x}$ and hyperparameters $\boldsymbol\theta$ through a series of nested approximations or a low-rank variational Bayes correction for the latent field \citep{Niekerk_2023_INLAv2, Niekerk_2024_lowrankVB}. This technique enables the computation of marginal likelihoods, standard goodness-of-fit metrics, cross-validation checks, and posterior predictive distributions \citep{Watanabe_2021_CrossValidationBayes, Watanabe_2023_InformationBayesInf, Liu_2025_LeaveGroupOut}.

Since INLA is based on GMRF principles, it offers inherent computational efficiency when modeling data with conditional independence. Indeed, if we have $n$ observations denoted as $\mathbf{y}=(y_1,\ldots,y_n)$, the likelihood of our data, assuming conditional independence, is described by the following expression:
\begin{equation}
p(\mathbf{y}|\boldsymbol\eta,\boldsymbol\theta)=\prod_{i=1}^n p(y_i \mid \eta_i,\boldsymbol\theta),
\label{eq:cond_indp_INLA}
\end{equation}
where the linear predictor is linked to the mean of the likelihood through a link function $g(\mu_i)=\eta_i$,  $\boldsymbol\eta$ is related to the latent field $\mathbf{x}$ through a projection matrix $\mathbf{A}$, such that $\boldsymbol\eta = \mathbf{A} \cdot \mathbf{x}$. The latent field $\mathbf{x}$ encompasses the vector of fixed effects $\boldsymbol\beta$ and the vector of random effects $\mathbf{f}$, i.e., $\mathbf{x} = (\boldsymbol\beta, \mathbf{f})$. The matrix $\mathbf{A}$ serves as the design matrix of the model, incorporating covariate values, factor structures and their levels, and weights associated with random effects (e.g., basis weights for splines). The hyperparameters are represented by $\boldsymbol\theta = (\boldsymbol\theta_1, \boldsymbol\theta_2)$, where $\boldsymbol\theta_1$ refers to the likelihood hyperparameters and $\boldsymbol\theta_2$ to the latent field hyperparameters. 

Assuming this conditional independence property and the structure of the latent field, the INLA methodology computes the posterior of the hyperparameters by examining the following functional relation in the modal configuration $\mathbf{x}^*(\boldsymbol\theta)$ of the Gaussian approximation for the conditional posterior of the latent field $\tilde{\pi}_G(\mathbf{x} \mid \mathbf{y}, \boldsymbol\theta)$:
\begin{equation}
    \tilde{\pi}(\boldsymbol\theta \mid \mathbf{y}) \propto \left.\frac{\pi(\mathbf{y}, \mathbf{x}, \boldsymbol\theta)}{\tilde{\pi}_G(\mathbf{x} \mid \mathbf{y}, \boldsymbol\theta)}\right\vert_{\mathbf{x} = \mathbf{x}^*(\boldsymbol\theta)}.
    \label{eq:marg_hyperpar_Std_INLA}
\end{equation}

The evaluation of this posterior distribution for the hyperparameters is performed at $K$ different support points, $\{\boldsymbol\theta^{k}\}^K_{k=1}$, based on a grid exploration or on a central composite design \citep[CCD,][]{Myers_2016_ResponseSurface}. The Gaussian approximation $\tilde{\pi}_G(\mathbf{x} \mid \mathbf{y}, \boldsymbol\theta)$ is calculated from a second-order Taylor expansion of the likelihood around $\mu^*(\boldsymbol\theta)$, the mode of $\pi(\mathbf{x} \mid \mathbf{y}, \boldsymbol\theta)$ \citep{Rue_2009_INLA, Niekerk_2023_INLAv2}:
\begin{equation}
\begin{array}{rcl}
\pi_{G}(\mathbf{x}\mid \mathbf{y}, \boldsymbol\theta) & \approx & \exp\big(\log\pi(\boldsymbol\mu(\boldsymbol\theta)\mid \mathbf{y},\boldsymbol\theta) + \nabla \left.\log\pi(\mathbf{x}\mid \mathbf{y}, \boldsymbol\theta)\right|_{\mathbf{x}=\boldsymbol\mu^*(\boldsymbol\theta)} \mathbf{x} \\[1.5mm]
 & & + \frac{1}{2}\mathbf{x}^\top \nabla^2\left.\log\pi(\mathbf{x}\mid \mathbf{y},\boldsymbol\theta)\right|_{\mathbf{x}=\boldsymbol\mu^*(\boldsymbol\theta)} \mathbf{x}) \; \big).
\end{array}
\label{eq:INLA_gaussian_approximation}
\end{equation}

This posterior approximation is obtained for each support point ${\boldsymbol{\theta}^k}$, and the mean is corrected through an iterative low-rank variational Bayes procedure \citep{Niekerk_2024_lowrankVB}. This yields a GMRF with a corrected mean for each support point ${\boldsymbol{\theta}^k}$:
\begin{equation}
\pi^{\prime}_G(\mathbf{x}\mid \mathbf{y}, \boldsymbol\theta^k) = \text{GMRF}(\boldsymbol\mu'(\boldsymbol\theta^k), \mathbf{Q}(\boldsymbol\theta^k)) ,
\end{equation}
where $\boldsymbol\mu^{\prime}(\boldsymbol{\theta})$ is the corrected mean, defined as $\boldsymbol\mu'(\boldsymbol{\theta}) = \boldsymbol\mu(\boldsymbol\theta) + \mathbf{M}\boldsymbol\lambda$. Here, $\mathbf{M}$ is a matrix that propagates the correction from $p$ pre-selected nodes to the rest of the latent field, and $\boldsymbol\lambda$ represents the explicit corrections at those $p$ nodes. The precision matrix is denoted by $\mathbf{Q}(\boldsymbol{\theta})$.

The GMRF of the latent field for each ${\boldsymbol{\theta}^k}$ is one of the outputs returned by the \texttt{inla} function in the \texttt{R-INLA} package, and is a key component for implementing our proposed distributed and recursive inference procedures.


After fixing the support points $\{\boldsymbol\theta^k\}_{k=1}^K$ the INLA approach computes the marginals of latent field using one of two approaches: (i) a nested Laplace approximation or a simplified Laplace approximation \citep{Rue_2009_INLA} or (ii) a low-rank Variational Bayes correction \citep{Niekerk_2024_lowrankVB} for the mean and marginal variances of the already computed conditional Gaussian approximation $\pi(\mathbf{x} \mid \mathbf{y}, \boldsymbol\theta)$. Finally, it performs a numerical integration over the hyperparameters to compute the latent field marginals:
\begin{equation}
    \pi(x_i \mid \mathbf{y}) = \sum_{k=1}^K \pi(x_i \mid \mathbf{y}, \boldsymbol\theta^k) \pi(\boldsymbol\theta^k \mid \mathbf{y}) \Delta_k.
    \label{eq:marginal_posterior_latentfield}
\end{equation}

A key aspect of this approach is the relationship between latent Gaussian fields (LGFs) and GMRFs. The latent field $\mathbf{x}$ follows a multivariate normal prior distribution with mean $\boldsymbol\mu$, usually equal to $\mathbf{0}$, and precision matrix $\mathbf{Q}(\boldsymbol\theta)$. This connection is crucial because, under the assumption that the LGF precision matrix is sparse, computational efficiency is significantly improved by enabling efficient sparse matrix operations \citep{Rue_2005_GMRF}.

\section{Distributed inference}
\label{sec:distributed_inference}


Distributed approaches are based on analyzing subsets of data separately and then recombining the results, where the subsets are defined either by a specific procedure or by a natural division of the data. In what follows, we present a new way to perform distributed inference within the INLA methodological framework \citep{Rue_2009_INLA, Niekerk_2023_INLAv2}, particularly in the context of the INLA version that implements a low-rank correction of the latent field \citep{Niekerk_2024_lowrankVB}. In the \emph{Supplementary Material}, we also describe different strategies to automatically construct the partitions.

In particular, this section is dedicated to a detailed presentation of the distributed approach for obtaining the marginal distributions of the latent field, $\pi(x_i \mid \mathbf{y})$, and the hyperparameters, $\pi(\theta_i \mid \mathbf{y})$, using the INLA methodology. This involves computing the conditional posterior distribution of the latent field, $\pi(\mathbf{x} \mid \boldsymbol\theta, \mathbf{y})$, presented in subsection 3.1, which is required to derive both the joint marginal distribution of the hyperparameters, $\pi(\boldsymbol\theta \mid \mathbf{y})$, presented in section 3.2, and the marginal distributions of the individual nodes in the latent field, $\pi(x_i \mid \mathbf{y})$, section 3.3. The marginal distribution for each hyperparameter, $\pi(\theta_i \mid \mathbf{y})$, is obtained by marginalizing the joint marginal posterior distribution of the hyperparameters, $\tilde{\pi}(\boldsymbol\theta \mid \mathbf{y})$, which is presented in subsection 3.4.

\subsection{Conditional posterior distribution of the latent field}
\label{sec:cond_post_distributed}

This distributed method relies on the conditional independence of the data given the latent field and the hyperparameters, as stated in Eq.~(\ref{eq:cond_indp_INLA}). This assumption allows us to partition or treat the data associated with the model in a distributed manner. The structure becomes more explicit when we write the conditional posterior distribution of the latent field, $\pi(\mathbf{x} \mid \mathbf{y}, \boldsymbol\theta)$, given the data and the hyperparameters. 

In the case of a general likelihood $\pi(\mathbf{y} \mid \mathbf{x}, \boldsymbol\theta)$ and a dataset $\mathbf{y}$, which can be split into $n$ subsets (or is naturally available in such a form), the posterior distribution of the latent field can be written by applying the conditional independence along the subsets $\mathbf{y}_i$:
\begin{equation}
\begin{array}{rcl}
    \pi(\mathbf{x} \mid \mathbf{y}, \boldsymbol\theta) & = & \displaystyle \pi(\mathbf{y} \mid \mathbf{x}, \boldsymbol\theta) \pi(\mathbf{x} \mid \boldsymbol\theta) \frac{\pi(\boldsymbol\theta)}{\pi(\boldsymbol\theta\mid \mathbf{y})\pi(\mathbf{y})} \\
     & \propto & \displaystyle \pi(\mathbf{y} \mid \mathbf{x}, \boldsymbol\theta)\pi(\mathbf{x} \mid \boldsymbol\theta) \\
     & \propto & \displaystyle \left[ \prod_{i = 1}^n\pi(\mathbf{y}_i \mid \mathbf{x}, \boldsymbol\theta) \right] \pi(\mathbf{x} \mid \boldsymbol\theta) \,,\\
\end{array}
\label{eq:cond_post_Gauss_like}
\end{equation}
where the conditional independence property is applied. From this step, it is possible to compute the conditional posterior distribution of the latent field in a distributed way. 

This conditional posterior can be computed analytically in the case of a Gaussian likelihood, since the INLA method is defined for Gaussian latent models and, therefore, conjugacy yields a closed-form expression for the posterior. Thus, in this section we consider two cases for computing the conditional posterior distribution of the latent field: one in which the likelihood is Gaussian, and another in which the likelihood is non-Gaussian.

\subsubsection{Conditional posterior with Gaussian likelihood}

In the Gaussian case, and assuming a linear predictor of the form $\boldsymbol\eta = \mathbf{A}\mathbf{x}$, where $\mathbf{A}$ is a projection matrix and $\mathbf{x}$ is a GMRF (typically with zero mean and precision matrix $\mathbf{Q}(\boldsymbol\theta_2)$), the expression can be made explicit as:
\begin{equation}
    \log\pi(\mathbf{x} \mid \mathbf{y}, \boldsymbol\theta) = C(\boldsymbol\theta, \mathbf{y}) -\frac{1}{2} \sum_{i=1}^n (\mathbf{y}_i - \mathbf{A}_i\mathbf{x})^\top\mathbf{W}(\boldsymbol\theta_1)(\mathbf{y}_i - \mathbf{A}_i\mathbf{x}) - \frac{1}{2} \mathbf{x}^\top \mathbf{Q}(\boldsymbol\theta_2)\mathbf{x} \,,
\end{equation}
where $C(\boldsymbol\theta, \mathbf{y})$ is a constant that depends on the hyperparameter vector $\boldsymbol\theta = \{\boldsymbol\theta_1, \boldsymbol\theta_2\}$ and incorporates contributions from the likelihood, the prior on the latent field $\pi(\mathbf{x}\mid\boldsymbol\theta_2)$, the hyperprior $\pi(\boldsymbol\theta)$, the marginal likelihood $\pi(\mathbf{y})$, and the marginal posterior $\pi(\boldsymbol\theta \mid \mathbf{y})$. If the conditional independence $\pi(\mathbf{y}_i, \mathbf{y}_j \mid \boldsymbol\eta, \boldsymbol\theta) = \pi(\mathbf{y}_i \mid \boldsymbol\eta, \boldsymbol\theta) \pi(\mathbf{y}_j \mid \boldsymbol\eta, \boldsymbol\theta)$ holds, then the matrix $\mathbf{W}(\boldsymbol\theta_1)$ is diagonal and can be written as $\mathbf{W}(\boldsymbol\theta_1) = \tau \mathbf{I}$, where $\tau$ is the marginal precision associated with the likelihood.

To make explicit the equivalence between the conditional posterior distribution from standard analysis and that from distributed inference, we note that the projection matrix $\mathbf{A}$ can be rewritten as: 
\begin{equation}
\mathbf{A} = \left(\begin{array}{l}
     \mathbf{A}_1  \\ \hline 
     \vdots \\ \hline
     \mathbf{A}_n
\end{array}\right) \quad \Rightarrow \quad \left\lbrace 
\begin{array}{l}
    \displaystyle
    \boldsymbol\eta = \mathbf{A}\mathbf{x} = \left(
    \begin{array}{c}
    \boldsymbol\eta_1  \\ \hline 
    \vdots \\ \hline
    \boldsymbol\eta_n
    \end{array}
    \right) = 
    \left(\begin{array}{l}
    \mathbf{A}_1 \mathbf{x}  \\ \hline 
    \vdots \\ \hline
    \mathbf{A}_n \mathbf{x}
\end{array}\right)  \\[10mm]
      \displaystyle\mathbf{x}^\top\mathbf{A}^\top \mathbf{A}\mathbf{x} = \boldsymbol\eta^\top \boldsymbol\eta = \sum_{i=1}^n \mathbf{x}^\top\mathbf{A}_i^\top\mathbf{A}_i\mathbf{x}
\end{array}\right. \qquad .
\end{equation}
Therefore, in the case of conditional independence, where $\mathbf{W}(\boldsymbol\theta_1) = \tau \mathbf{I}$,  the log-posterior conditional distribution can be written as:
\begin{equation}
    \resizebox{.9\textwidth}{!}{
    $
    \begin{array}{rcl}
        \log\pi(\mathbf{x} \mid \mathbf{y}, \boldsymbol\theta) & = &  \displaystyle C(\boldsymbol\theta, \mathbf{y}) -\frac{1}{2} \left(\tau\cdot\mathbf{x}^\top\mathbf{A}^\top\mathbf{A}\mathbf{x} - 2\cdot\tau\cdot \mathbf{y}^\top\mathbf{A}\mathbf{x} \right) - \frac{1}{2}\mathbf{x}^\top\mathbf{Q}(\boldsymbol\theta_2)\mathbf{x} \\
        & = & \displaystyle C(\boldsymbol\theta, \mathbf{y}) -\frac{1}{2} \sum_{i=1}^n \left(\tau\cdot\mathbf{x}^\top\mathbf{A}_i^\top\mathbf{A}_i\mathbf{x} - 2\cdot\tau\cdot \mathbf{y}^\top_i\mathbf{A}_i\mathbf{x} \right) - \frac{1}{2}\mathbf{x}^\top\mathbf{Q}(\boldsymbol\theta_2)\mathbf{x} \\
        & = & \displaystyle C(\boldsymbol\theta, \mathbf{y}) -\frac{1}{2} \sum_{i=1}^n \left(\tau\cdot\mathbf{x}^\top\mathbf{A}_i^\top\mathbf{A}_i\mathbf{x} - 2\cdot\tau\cdot \mathbf{y}^\top_i\mathbf{A}_i\mathbf{x} + \frac{1}{n}\mathbf{x}^\top\mathbf{Q}(\boldsymbol\theta_2)\mathbf{x}\right) \\
        & = & \displaystyle C^*(\boldsymbol\theta, \mathbf{y}) + \sum_{i=1}^n \left(\log\pi(\mathbf{y}_i \mid \mathbf{A}_i \mathbf{x}, \boldsymbol\theta) + \frac{1}{n}\log\pi(\mathbf{x}\mid \boldsymbol\theta_2)\right) \\
        & = & \displaystyle C^\prime(\boldsymbol\theta, \mathbf{y}) + \sum_{i=1}^n \log\pi(\mathbf{x} \mid \mathbf{y}_i, \boldsymbol\theta)
    \end{array}
    $}
\label{eq:conditional_distributed_extended}
\end{equation}
where $C(\boldsymbol\theta, \mathbf{y})$, $C^*(\boldsymbol\theta, \mathbf{y})$ and $C^\prime(\boldsymbol\theta, \mathbf{y})$ are constants ensuring the equivalence of the different formulations of $\log\pi(\mathbf{x}\mid\mathbf{y}, \boldsymbol\theta)$. These constants can be ignored since the posterior is Gaussian, and given the mean and the precision matrix, we can immediately recover the analytical form of the posterior distribution without the need to compute the constants. Furthermore, when computing the posterior distribution associated with each partition $\log\pi(\mathbf{x} \mid \mathbf{y}_i, \boldsymbol\theta)$ one can use a prior weighted identically across partitions---as done above---or a prior weighted according to the number of observations in each partition:
\begin{equation}
    \log\pi(\mathbf{x} \mid \mathbf{y}, \boldsymbol\theta) = \displaystyle C^*(\boldsymbol\theta, \mathbf{y}) + \sum_{i=1}^n \left(\log\pi(\mathbf{y}_i \mid \mathbf{A}_i \mathbf{x}, \boldsymbol\theta) + w_i\log\pi(\mathbf{x}\mid \boldsymbol\theta_2)\right),
\end{equation}
where the weights can be defined as $w_i = (|\mathbf{y}| - |\mathbf{y}_i|)/((n-1)\cdot|\mathbf{y}|)$ such that partitions with fewer observations are assigned priors that are weighted more heavily, thus remaining closer to the original prior. Any alternative weighting scheme that satisfies $w_i>0$ $(\forall i)$ and $\sum_{i=1}^nw_i=1$ can, in principle, also be used. 

An alternative to weighting the prior distribution is to use the unweighted prior and modify the posterior distribution as follows:
\begin{equation}
    \resizebox{.9\textwidth}{!}{$
    \log\pi(\mathbf{x} \mid \mathbf{y}, \boldsymbol\theta) = \displaystyle C(\boldsymbol\theta, \mathbf{y}) + \sum_{i=1}^n \left(\log\pi(\mathbf{y}_i \mid \mathbf{A}_i \mathbf{x}, \boldsymbol\theta) + \log\pi(\mathbf{x}\mid \boldsymbol\theta_2)\right) - (n- 1)\cdot \log\pi(\mathbf{x}\mid\boldsymbol\theta_2),
    $}
\end{equation}
which is equivalent to compute the posterior as
$$\pi(\mathbf{x \mid \mathbf{y}, \boldsymbol\theta}) \propto \pi^*(\mathbf{x}\mid\mathbf{y}, \boldsymbol\theta)/\pi(\mathbf{x} \mid \boldsymbol\theta)^{n-1},$$ 
where $\pi^*(\mathbf{x} \mid \mathbf{y}, \boldsymbol\theta)$ is the posterior computed in a distributed fashion, using the unweighted prior in each partition.

This allows the conditional posterior distribution of the latent field in each partition, $\pi(\mathbf{x} \mid \mathbf{y}_i, \boldsymbol\theta)$, to be computed independently, either in parallel on different machines or servers, or in parallel on a single machine. The latter may be particularly useful, as we will discuss later, when partitioning the data is done through the partitioning of the latent field, generally reducing the computational cost of computing each of these posterior distributions. The distributed conditional posterior for the case of a Gaussian likelihood and under conditional independence of the observations---as discussed earlier---is analytically identical to the posterior obtained without data partitioning. However, this identity no longer holds in the case of non-Gaussian likelihoods, and the distributed approach becomes an approximation.

\subsubsection{Conditional posterior with non-Gaussian likelihood}

In general, for non-Gaussian likelihoods, the conditional posterior for each partition is approximated via a second-order Taylor expansion of the log-likelihood around the mode of the true conditional posterior $\pi(\mathbf{x} \mid \mathbf{y}_i, \boldsymbol\theta)$. The Gaussian approximation uses the mode $\boldsymbol\mu_i(\boldsymbol\theta)$ for the $i$-th partition, computed as the minimum of the negative log conditional posterior
$$
\boldsymbol\mu_i(\boldsymbol\theta) = \underset{\mathbf{x}}{\argmin}\left\lbrace -\log\pi(\mathbf{x} \mid \mathbf{y}_i, \boldsymbol\theta) \right\rbrace \,,
$$
and the precision matrix at the mode. This precision matrix is computed by the Hessian of the negative log-conditional posterior evaluated at $\boldsymbol\mu_i(\boldsymbol\theta)$:
$$
\mathbf{Q}_i(\boldsymbol\theta) = -\left. \mathbf{H}(\log\pi(\mathbf{x} \mid \mathbf{y}_i, \boldsymbol\theta))\right\vert_{x = \boldsymbol\mu_i(\boldsymbol\theta)} = \mathbf{Q}(\boldsymbol\theta) + \mathbf{A}_i^\top\mathbf{D}_i\mathbf{A}_i,
$$
where $\mathbf{Q}(\boldsymbol\theta)$ is the prior precision matrix for the latent field, $\mathbf{A}_i$ is the projection matrix for the linear predictor, and $\mathbf{D}_i$ is a diagonal matrix computed from the second derivatives of the likelihood. 

Hence, the conditional posterior distribution is approximated by a Gaussian distribution $\pi_G(\mathbf{x} \mid \mathbf{y}_i, \boldsymbol\theta)$ for each partition, expressed in canonical form:
\begin{equation}
    \pi_G(\mathbf{x} \mid \mathbf{y}_i, \boldsymbol\theta) \propto \exp\left[ -\frac{1}{2}\mathbf{x}^\top\left(w_i\mathbf{Q}(\boldsymbol\theta) + \mathbf{A}_i^\top\mathbf{D}_i\mathbf{A}_i \right) \mathbf{x} - \mathbf{b}_i^\top \mathbf{A}_i \mathbf{x}  \right]
\end{equation}
where $(w_i\mathbf{Q}(\boldsymbol\theta) + \mathbf{A}_i^\top\mathbf{D}_i\mathbf{A}_i)$ is the precision matrix of the approximated posterior and the prior precision is weighted by $w_i$, as explained earlier. The mean is computed as $\mathbf{b}_i = w_i\mathbf{Q}_i^{-1}(\boldsymbol\theta)\boldsymbol\mu_i$. This gives a Gaussian approximation for the conditional posterior of the latent field for each partition $\mathbf{y_i}$: 
$$
\pi_G(\mathbf{x} \mid \mathbf{y}_i, \boldsymbol\theta)= \text{N}(\boldsymbol\mu_i(\boldsymbol\theta), \mathbf{Q}_i(\boldsymbol\theta)).
$$

Therefore, to obtain the Gaussian approximation for the full dataset, one could apply any of the strategies discussed for the Gaussian case. However, with non-Gaussian likelihoods, the identity between the full and distributed conditional posteriors no longer holds. The posterior for the complete dataset is:
$$
\pi_G(\mathbf{x} \mid \mathbf{y}, \boldsymbol\theta) \propto \exp\left[ -\frac{1}{2}\mathbf{x}^\top\left(\mathbf{Q}(\boldsymbol\theta) + \mathbf{A}^\top\mathbf{D}\mathbf{A} \right) \mathbf{x} - \mathbf{b}^\top \mathbf{A} \mathbf{x}  \right]\; ,
$$
which implies that the components $\mathbf{D}$ and $\mathbf{b}$ will differ across partitions. 

Therefore, the identity used in the Gaussian case no longer holds, although we may still use this product as an approximation:
\begin{equation}    
    \pi_G(\mathbf{x} \mid \mathbf{y}, \boldsymbol\theta) \approx \pi^*_G(\mathbf{x} \mid \mathbf{y}, \boldsymbol\theta) \approx C(\boldsymbol\theta, \mathbf{y})\cdot \prod_{i=1}^n \pi_G(\mathbf{x} \mid \mathbf{y}_i, \boldsymbol\theta),
    \label{eq:distributed_conditional_posterior}
\end{equation}
where the quality of this approximation depends on the quality of each individual approximation $\pi_G(\mathbf{x} \mid \mathbf{y}, \boldsymbol\theta)$ across all partitions. In other words, the approximation $\pi^*_G(\mathbf{x} \mid \mathbf{y}, \boldsymbol\theta)$ accumulates the errors of the per-partition approximations, whereas the global approximation $\pi_G(\mathbf{x} \mid \mathbf{y}, \boldsymbol\theta)$ can achieve greater accuracy, as it uses all available data to better identify the mode and the curvature at the mode.

The error made in the Gaussian approximation in the INLA methodology is discussed by \cite{Martino_2007_INLAv0}, where it is shown that the error in the approximation $\pi_G(\mathbf{x} \mid \boldsymbol\theta, \mathbf{y})$ arises primarily from mode location and skewness. Therefore, \cite{Niekerk_2023_INLAv2} propose correcting the mean of the Gaussian using a low-rank variational Bayes approach while keeping the variance unchanged, instead of relying in more computational demanding strategies as Laplace or simplified Laplace \citep{Rue_2009_INLA}.



\subsection{Joint marginal posterior distribution of the hyperparameters}

The next step, following the inferential structure of INLA presented in Section 2, is to obtain the marginal posterior distribution of the hyperparameters $\pi(\boldsymbol\theta \mid \mathbf{y})$, which is generally obtained approximately using the functional form evaluated at the modal configuration of the conditional posterior of the latent field, as shown in Eq.~(\ref{eq:marg_hyperpar_Std_INLA}):
$$
\tilde{\pi}(\boldsymbol\theta \mid \mathbf{y}) \propto \left. \frac{\pi(\mathbf{y} \mid \mathbf{x}, \boldsymbol\theta) \pi(\mathbf{x}\mid\boldsymbol\theta)\pi(\boldsymbol\theta)}{\pi_G(\mathbf{x} \mid \mathbf{y},\boldsymbol\theta)} \right|_{\mathbf{x=\mathbf{x}^*(\boldsymbol\theta)}},
$$
where $\pi(\mathbf{y} \mid \mathbf{x},\boldsymbol\theta)$ is the likelihood, $\pi(\mathbf{x} \mid \boldsymbol\theta)$ is the prior distribution for the latent field, $\pi(\boldsymbol\theta)$ is the prior for the hyperparameters, and $\pi_G(\mathbf{x} \mid \mathbf{y}, \boldsymbol\theta)$ is the conditional posterior distribution previosly discussed, which is exact in the case of a Gaussian likelihood and approximated by a Gaussian distribution for non-Gaussian likelihoods. However, regardless of whether this is exact or a Gaussian approximation, evaluating it at the modal configuration of the conditional posterior distribution of the latent field $\pi(\mathbf{x} \mid \mathbf{y}, \boldsymbol\theta)$ leads to the following expression:
\begin{equation}
    \tilde{\pi}(\boldsymbol\theta \mid \mathbf{y}) \propto \left.\frac{\pi(\mathbf{y}\mid\boldsymbol\theta)|\mathbf{Q}(\theta)|\exp\left(-\frac{1}{2}\mathbf{x}^\top\mathbf{Q}(\theta)\mathbf{x}\right)}{|\mathbf{Q}_{\mathbf{x}\mid\mathbf{y},\boldsymbol\theta}(\boldsymbol\theta)|}\right|_{\mathbf{x}=\mathbf{x}^*(\boldsymbol\theta)} \pi(\boldsymbol\theta),
    \label{eq:marginal_posterior_hyper_Gap}
\end{equation}
where $|\mathbf{Q}(\theta)|$ is the determinant of the precision matrix of the prior distribution of the latent field, and $|\mathbf{Q}_{\mathbf{x}\mid\mathbf{y},\boldsymbol\theta}(\boldsymbol\theta)|$ is the determinant of the precision matrix of the conditional posterior distribution of the latent field.

An important issue at this step arises when constraints are imposed on elements of the latent field, $\mathbf{C} \mathbf{x} = \mathbf{e}$. When such constraints are introduced, the precision matrix of the posterior distribution can be computed using \textit{conditioning by kriging} \citep{Rue_2009_INLA}, such that the precision matrix with constraints, denoted $\mathbf{Q}_c$, is given by:
\begin{equation}
    \mathbf{Q}_c^{-1} = \mathbf{Q}^{-1} - \mathbf{Q}^{-1} \mathbf{C}^\top (\mathbf{C} \mathbf{Q}^{-1} \mathbf{C}^\top)^{-1} \mathbf{C}\mathbf{Q}^{-1},
    \label{eq:conditioning_kriging}
\end{equation}
and the determinant is computed over this precision matrix that incorporates the constraints. In \cite{Fattah_2022_ABInteraction}, an alternative approach is proposed for the case of space-time models with interaction \citep{KnorrHeld_SpaceTime_2000}, to compute the covariance matrix $\boldsymbol\Sigma_c = \mathbf{Q}_c^{-1}$ associated with the conditional posterior distribution of the latent field, where the constraints are embedded by design in the null space structure of the prior distribution of the space-time effect. This approach leverages the Sherman–Morrison–Woodbury identity (also known as the Woodbury formula).

To compute the joint marginal posterior distribution of the hyperparameters in the distributed framework proposed in this paper, we can consider different strategies. One option is to compute the marginal distribution as stated in Eq.~\eqref{eq:distributed_conditional_posterior}. This strategy requires evaluating the marginal posterior of the hyperparameters on a single machine with access to the full dataset, which is not feasible in privacy-sensitive settings. Alternatively, we can propose computing the marginal posterior leveraging a distributed framework, denoted by $\tilde{\pi}_d(\boldsymbol\theta \mid \mathbf{y})$, where the posterior is computed separately for each dataset and then combined across the $n$ different subsets $\mathbf{y}= \{\mathbf{y}_1, \dots, \mathbf{y}_n\}$:
\begin{equation}
    \tilde{\pi}_d(\boldsymbol\theta \mid \mathbf{y}) \propto \prod_{i=1}^n \tilde{\pi}_d(\boldsymbol\theta \mid \mathbf{y}_i) = \prod_{i=1}^n \left.\frac{\pi(\mathbf{y}_i \mid \mathbf{x}, \boldsymbol\theta)\pi(\mathbf{x} \mid \boldsymbol\theta)^{w_i} \pi(\boldsymbol\theta)^{w_i}}{\pi_G(\mathbf{x} \mid \boldsymbol\theta, \mathbf{y}_i)} \right|_{\mathbf{x=\mathbf{x}_i^*(\boldsymbol\theta)}},
    \label{eq:marginal_posterior_hyper_distributed}
\end{equation}
where the weights $w_i$ satisfy $\sum_{i=1}^n w_i = 1$ and $w_i > 0$ $(\forall i)$, as defined in the previous section regarding the conditional posterior distribution of the latent field. Furthermore, $\pi_G(\mathbf{x} \mid \boldsymbol\theta, \mathbf{y}_i)$ evaluated at the mode $\mathbf{x}=\mathbf{x}_i^*(\boldsymbol\theta)$ reduces to
$$
\pi_G(\mathbf{x}=\mathbf{x}_i^*(\boldsymbol\theta) \mid \boldsymbol\theta, \mathbf{y}_i) \propto |\mathbf{Q}^c_{\mathbf{x}\mid\boldsymbol\theta,\mathbf{y}_i}(\boldsymbol\theta)|,
$$
where it is assumed that this posterior has been computed by incorporating the relevant constraints. 

In summary, the two possible strategies to compute the joint marginal posterior of the hyperparamters are: (1) obtain the conditional posterior distribution via Eq.(\ref{eq:distributed_conditional_posterior}) and apply Eq.(\ref{eq:conditioning_kriging}) to compute the marginal posterior via Eq.(\ref{eq:marginal_posterior_hyper_Gap}), or (2) use the distributed approach described in Eq.(\ref{eq:marginal_posterior_hyper_distributed}) and apply Eq.~(\ref{eq:conditioning_kriging}) separately in each partition.

\subsection{Marginal posterior distributions of the latent field}

One of the key elements in the INLA framework is to find the mode of the marginal posterior distribution of the hyperparameters using optimization techniques based on finite differences, since this distribution is not analytically differentiable \citep{Rue_2009_INLA}. From this mode, it is then possible to perform an exploration strategy (e.g. grid, central composite design) of the marginal posterior and compute both the marginal distributions of the nodes of the latent field and the marginal distributions of each hyperparameter.

To construct the integration scheme---particularly the Central Composite Design (CCD)---we can start from the distributed expression of the marginal posterior of the hyperparameters in Eq.~(\ref{eq:marginal_posterior_hyper_distributed}); based on this expression, we can compute the marginal distributions of the nodes of the latent field, $\pi(x_i \mid \mathbf{y})$, as well as the marginal distributions of the individual hyperparameters, $\pi(\theta_i \mid \mathbf{y})$.

In the distributed approach, instead of estimating the mode of $\tilde{\pi}(\boldsymbol\theta \mid \mathbf{y})$ directly, we can construct a Gaussian approximation of the mode of $\tilde{\pi}(\boldsymbol\theta \mid \mathbf{y})$ based on the Gaussian approximations at the modes of the individual posteriors $\tilde{\pi}d(\boldsymbol\theta \mid \mathbf{y}i)$. Thus, the Gaussian approximation of $\tilde{\pi}(\boldsymbol\theta \mid \mathbf{y})$ can be defined as:
\begin{equation}
    \tilde{\pi}_G(\boldsymbol\theta \mid \mathbf{y}) = \text{N}(\boldsymbol\mu_{\boldsymbol\theta^*}, \boldsymbol\Sigma_{\boldsymbol\theta^*}),
\end{equation}
where $\boldsymbol\mu_{\boldsymbol\theta^*} = \boldsymbol\theta^*$ is the mode of $\tilde{\pi}(\boldsymbol\theta \mid \mathbf{y})$, and $\boldsymbol\Sigma_{\boldsymbol\theta^*}$ is the covariance matrix evaluated at the mode $\boldsymbol\mu_{\boldsymbol\theta^*}$ of the marginal posterior distribution.

The mode is obtained by minimizing $-\log(\tilde{\pi}(\boldsymbol\theta \mid \mathbf{y}))$ using finite difference methods:
$$
\boldsymbol\mu_{\boldsymbol\theta^*} = \boldsymbol\theta^* = \underset{\boldsymbol\theta}{\argmin}\{-\log(\tilde{\pi}(\boldsymbol\theta \mid \mathbf{y}))\}
$$
and the covariance matrix is computed as the inverse of the Hessian matrix of $-\log(\tilde{\pi}(\boldsymbol\theta \mid \mathbf{y}))$, that is, $\boldsymbol\Sigma^{-1}_{\boldsymbol\theta^*} = -\mathbf{H}_{\boldsymbol\theta^*}(\log\tilde{\pi}(\boldsymbol\theta \mid \mathbf{y}))$, evaluated at the mode:
$$
\mathbf{H}_{\boldsymbol\theta^*}(\log\tilde{\pi}(\boldsymbol\theta \mid \mathbf{y})) = \left.\left(\begin{array}{ccc}
    \frac{\partial^2}{\partial^2\theta_1} \log(\tilde{\pi}(\boldsymbol\theta \mid \mathbf{y})) &  \cdots & \frac{\partial^2}{\partial\theta_1\partial\theta_m}\log(\tilde{\pi}(\boldsymbol\theta \mid \mathbf{y})) \\
    \vdots & \ddots & \vdots \\
     \frac{\partial^2}{\partial\theta_m\partial\theta_1} \log(\tilde{\pi}(\boldsymbol\theta \mid \mathbf{y})) & \cdots & \frac{\partial^2}{\partial^2\theta_m} \log(\tilde{\pi}(\boldsymbol\theta \mid \mathbf{y}))
\end{array} \right) \right\vert_{\boldsymbol\theta^*} ,
$$
where these second-order derivatives are also estimated by finite differences, since the expression for the marginal posterior of the hyperparameters, as defined in Eq.~(\ref{eq:marg_hyperpar_Std_INLA}), is not analytically tractable.

This Gaussian approximation at the modal configuration of $\tilde{\pi}(\boldsymbol\theta \mid \mathbf{y})$ serves as the starting point for constructing a global approximation using the Gaussian approximations associated with the modes of the marginal posteriors from each partition, $\tilde{\pi}_d(\boldsymbol\theta \mid \mathbf{y}_i)$:
\begin{equation}
    \tilde{\pi}_G(\boldsymbol\theta \mid \mathbf{y}) \approx \prod_{i=1}^n \pi_G(\boldsymbol\theta \mid \mathbf{y}_i), 
    \label{eq:dist_hyper_to_CCD_1}
\end{equation}
where the precision matrix associated with $\tilde{\pi}_G(\boldsymbol\theta \mid \mathbf{y})$ is computed as the sum of the precision matrices evaluated at the modal configuration of the marginal posterior of the hyperparameters for each partition. That is, the precision matrix is approximated as:
\begin{equation}
    \mathbf{H}_{\boldsymbol\theta^*} \approx \sum_{i=1}^n \mathbf{H}_{\boldsymbol\theta_i^*}(\log\tilde{\pi}_d(\boldsymbol\theta \mid \mathbf{y}_i)),
    \label{eq:dist_hyper_to_CCD_2}
\end{equation}
where $\mathbf{H}_{\boldsymbol\theta_i^*}(\log\tilde{\pi}_d(\boldsymbol\theta \mid \mathbf{y}_i))$ is the Hessian of $\log\tilde{\pi}_d(\boldsymbol\theta \mid \mathbf{y}_i)$ evaluated at the corresponding mode $\boldsymbol\theta^*_i$, which is specific to each data partition.

The mean for each partition is given by $\boldsymbol\mu_{\boldsymbol\theta^*_i} = \boldsymbol\theta_i^*$, as in previous cases where the Gaussian approximation has been applied. Therefore, the mean of the global Gaussian approximation $\tilde{\pi}_G(\boldsymbol\theta \mid \mathbf{y})$ is approximated by:
\begin{equation}
    \boldsymbol\mu_{\boldsymbol\theta^*} \approx \mathbf{H}^{-1}_{\boldsymbol\theta^*}\sum_{i=1}^n \mathbf{H}_{\boldsymbol\theta_i^*}\boldsymbol\mu_{\boldsymbol\theta_i^*}.
    \label{eq:dist_hyper_to_CCD_3}
\end{equation}

Consequently, the Gaussian approximation at the modal configuration of the hyperparameters is given by:
\begin{equation}
    \tilde{\pi}_G(\boldsymbol\theta \mid \mathbf{y}) \approx \text{N}\left(\left[\sum_{i=1}^n\mathbf{H}_{\boldsymbol\theta_i^*}\right]^{-1}\sum_{i=1}^n \mathbf{H}_{\boldsymbol\theta_i^*}\boldsymbol\mu_{\boldsymbol\theta_i^*}, \sum_{i=1}^n \mathbf{H}_{\boldsymbol\theta_i^*} \right)
    \label{eq:dist_hyper_to_CCD_4}
\end{equation}
expressed in terms of the mean and precision matrix. Based on this Gaussian approximation, we can construct the CCD and compute the weights $\Delta_k$ related to the integration as explained in \cite{Rue_2009_INLA}.

Once the integration scheme has been defined, with its associated support points $\{\boldsymbol\theta^k\}_{k=1}^K$, we can compute the marginal distribution of the latent field parameters using Eq.~(\ref{eq:marginal_posterior_latentfield}), which incorporates the conditional posterior of the latent field, the marginal of the hyperparameters, and the weights from the integration scheme defined in the distributed approach:

\begin{equation}
    \tilde{\pi}_d(x_i \mid \mathbf{y}) = \sum_{i=1}^{n} \tilde{\pi}_G(x_i \mid \boldsymbol\theta^k, \mathbf{y}) \tilde{\pi}_d(\boldsymbol\theta^k \mid \mathbf{y}) \Delta_k
    \label{eq:distr_marginal_nodes}
\end{equation}
where $\tilde{\pi}_G(x_i \mid \boldsymbol\theta^k, \mathbf{y})$ denotes the marginal distribution of node $x_i$ from the conditional posterior of the latent field. 

Note that at this stage we can apply the low-rank correction of the mean via Variational Bayes \citep{Niekerk_2024_lowrankVB} to the conditional posterior distribution $\tilde{\pi}_G(\mathbf{x} \mid \boldsymbol\theta, \mathbf{y})$, obtained after aggregating all information from the different data subsets. Alternatively, we could apply the low-rank correction to the conditional posterior distribution of each subset, i.e., $\tilde{\pi}_G(\mathbf{x} \mid \boldsymbol\theta, \mathbf{y}_i)$, before merging the information. The choice between these two approaches depends on whether the full dataset is centrally accessible, or if privacy constraints restrict simultaneous data access. Furthermore, the constraints imposed on the latent field have already been incorporated into its conditional posterior distribution during the computation of the hyperparameter marginal posteriors.

\subsection{Marginal posterior distributions of the hyperparameters}

The marginal posterior distributions of each hyperparameter, $\tilde{\pi}(\theta_i \mid \mathbf{y})$, can be computed by following different strategies, such as those presented in \cite{Rue_2009_INLA} and \cite{Martins_2013_INLAsoftware} along with a consensus approach \citep{Scott_2016_ConsensusMonteCarlo}, which can be summarized into the following five approaches: (i) interpolation, (ii) asymmetric Gaussian interpolation, (iii) Laplace approximation, (iv) the numerical integration-free algorithm and, (v) consensus Monte Carlo.

The first strategy---interpolation---requires a grid exploration to obtain a sufficient number of evaluation points for interpolation. In particular, one can use the grid exploration described in \cite{Rue_2009_INLA} and \cite{Martins_2013_INLAsoftware}, and, based on the set of support points $\{\boldsymbol\theta^k\}_{k=1}^K$ where the posterior density $\{\tilde{\pi}(\boldsymbol\theta^k \mid \mathbf{y})\}_{k=1}^K$ has been evaluated, construct an interpolant $I(\boldsymbol\theta \mid \mathbf{y})$ such that the marginalization over each dimension of the hyperparameters can be computed numerically:
$$
\tilde{\pi}(\theta_j \mid \mathbf{y}) \approx \int_{\boldsymbol\theta_{-j}\in \boldsymbol\Theta_{-j}} I(\boldsymbol\theta \mid \mathbf{y}) d\boldsymbol\theta_{-j},
$$
where the interpolant $I(\boldsymbol\theta \mid \mathbf{y})$ can be selected from among a wide variety of available interpolation methods---for example, inverse distance weighting interpolation (and its variants), nearest-neighbour interpolation, radial basis function interpolation, kernel interpolation, or barycentric interpolation. An alternative grid strategy for exploring the posterior distribution of the hyperparameters involves using Korobov lattices and low discrepancy sequences, as described in \cite{Brown_2021_HyperKorobovLDS}.

The second strategy considers that, under a rescaling and rotation of the internal parameterization of the hyperparameters, given by $\boldsymbol\theta(\mathbf{z}) = \boldsymbol\theta^* + \mathbf{U}\boldsymbol\Lambda^{1/2}\mathbf{z}$—where $\mathbf{U}$ and $\boldsymbol\Lambda$ are, respectively, the matrix of eigenvectors and the diagonal matrix of eigenvalues from the eigendecomposition of $\mathbf{H}_{\boldsymbol\theta^*}$---we can employ an \textit{asymmetric Gaussian approximation} over this reparameterization $\mathbf{z}(\boldsymbol\theta)$. First, we define the function
\begin{equation}
    f(\boldsymbol\theta) = \prod_{j=1}^J f_j(z_j(\boldsymbol\theta))
    \label{eq:hyperpar_marg_assymetricGauss}
\end{equation} 
 where each component function $f_j(z_j(\boldsymbol\theta))$ is defined as:
\begin{equation}
f_j(z_j(\boldsymbol\theta)) \propto \left\lbrace\begin{array}{c}
      \exp\left(-\frac{1}{2(\sigma_j^-)^2}z_j^2\right) \quad \text{if} \quad z_j<0 \\
      \exp\left(-\frac{1}{2(\sigma_j^+)^2}z_j^2\right) \quad \text{if} \quad z_j\geq 0
\end{array} \right. \qquad.
\end{equation}

These functions $f_j(z_j(\boldsymbol\theta))$ can be interpreted as a Gaussian approximation capable of capturing the asymmetry of $\tilde{\pi}(\boldsymbol\theta \mid \mathbf{y})$. In each dimension of $\mathbf{z}(\boldsymbol\theta) = (z_1(\boldsymbol\theta), ..., z_J(\boldsymbol\theta))$, a piecewise-defined function is used, with a Gaussian kernel on each side. Each side of the axis $z_j(\boldsymbol\theta)$ has a different variance, allowing for a more flexible representation of the asymmetry of $\tilde{\pi}(\boldsymbol\theta \mid \mathbf{y})$ depending on the direction with respect to the coordinate center in the transformed space $\mathbf{z}(\boldsymbol\theta)$. The scaling parameters $(\sigma_j^-, \sigma_j^+)$ can be estimated during the optimization process used to find the mode of Eq.(\ref{eq:marg_hyperpar_Std_INLA}). Once these scaling parameters are obtained, the marginal posteriors $\tilde{\pi}(\theta_j \mid \mathbf{y})$ can be computed via numerical integration of Eq.(\ref{eq:hyperpar_marg_assymetricGauss}).

Nevertheless, these numerical integration approaches require evaluating Eq.(\ref{eq:marg_hyperpar_Std_INLA})---or an approximation thereof, as in the second strategy---a considerable number of times, making both strategies computationally expensive, particularly the first. An alternative to these two interpolation-based strategies would be to replace the grid-based exploration, which typically implies a sequential search strategy, with direct sampling from the Gaussian approximation at the mode. This would allow for parallel evaluation of Eq.(\ref{eq:marg_hyperpar_Std_INLA}), thus significantly improving computational efficiency.

The third strategy consists in applying another Laplace approximation to compute the marginal distributions:
\begin{equation}
    \tilde{\pi}(\theta_j \mid \mathbf{y}) = \left.\frac{\tilde{\pi}(\boldsymbol\theta \mid \mathbf{y})}{\tilde{\pi}_{G}(\boldsymbol\theta_{-j} \mid \theta_j, \mathbf{y})}\right|_{\boldsymbol\theta_{-j} = \boldsymbol\theta^*_{-j}},
\end{equation}
where $\boldsymbol\theta_{-j}$ is the modal configuration of $\tilde{\pi}(\boldsymbol\theta_{-j} \mid \theta_j, \mathbf{y})$. While this approach can yield accurate results, it requires finding the maximum of the $(J-1)$-dimensional function $\pi(\boldsymbol\theta_{-j} \mid \theta_j, \mathbf{y})$ for each value of $\theta_j$, which does not scale efficiently with the dimensionality $J$ of hyperparameter space.

The fourth strategy, the \textit{numerical integration free algorithm}, is the default one in the \texttt{R-INLA} software, and consists in assuming the following approximation for the marginals
\begin{equation}
\tilde{\pi}(\theta_j \mid \mathbf{y}) \propto \left\lbrace\begin{array}{c}
      \exp\left(-\frac{1}{2(\sigma_j^-)^2}\theta_j^2\right) \quad \text{if} \quad \theta_j<0 \\
      \exp\left(-\frac{1}{2(\sigma_j^+)^2}\theta_j^2\right) \quad \text{if} \quad \theta_j\geq 0
\end{array} \right. \qquad.   
\label{eq:dist_marg_hyper_free}
\end{equation}
where, as in the asymmetric Gaussian interpolation, we need to compute the scaling parameters $(\sigma_j^-, \sigma_j^+)$, but now without interpolation. To accomplish this, we can leverage the following lemma from \cite{Rue_2009_INLA}:
\begin{equation}
-\frac{1}{2} (\theta_j, \mathbb{E}(\boldsymbol\theta_{-j} \mid \theta_j)) \boldsymbol\Sigma^{-1} \left(\begin{array}{c} \theta_j \\ \mathbb{E}(\boldsymbol\theta_{-j} \mid \theta_j)\end{array}\right) = -\frac{1}{2} \frac{\theta_j^2}{\Sigma_{jj}}    ,
\label{eq:lemma_marg_hyper}
\end{equation}
which states that the values of the joint density of $\boldsymbol\theta$, considered as a function of $\theta_j$ and $\boldsymbol\theta_{-j}$, evaluated at the conditional mean $\mathbb{E}(\boldsymbol\theta_{-j} \mid \theta_j)$, behaves as the marginal distribution of $\theta_i$. Using this result, we can explore both sides of $\theta_j$ for each hyperparameter and evaluate, at least, three points to estimate the curvature, i.e. the scaling parameters $(\sigma_j^-, \sigma_j^+)$.

This approach assumes that the joint marginal distribution of the hyperparameters behaves like its Gaussian approximation around the mode, which allows us to compute the conditional expectation and apply the previously stated lemma. This, in turn, implies that at each support point in the hyperparameter space, the Hessian with respect to the hyperparameters must be computed. In the distributed setting, this requires calculating the Hessian separately for each partition and then applying Eq.~\eqref{eq:dist_hyper_to_CCD_2} to obtain the overall Hessian at each support point. 

Finally, the fifth strategy is based on consensus Monte Carlo \citep{Scott_2016_ConsensusMonteCarlo}. This approach leverages the fact that we can sample from the approximate joint marginal posterior distribution of the hyperparameters for each partition, denoted as $\tilde{\pi}_d(\boldsymbol\theta \mid \mathbf{y}_i)$. Specifically, by computing the asymmetric Gaussian approximation via the numerical integration-free algorithm, we can readily draw samples from this approximate distribution. Consequently, given the local approximations $\tilde{\pi}_d(\boldsymbol\theta \mid \mathbf{y}_i)$, the global marginal distribution for the hyperparameters in the distributed framework can be computed as the product:
\begin{equation}
    \tilde{\pi}_d(\boldsymbol\theta \mid \mathbf{y}) = \prod_{i=1}^n \tilde{\pi}_d(\boldsymbol\theta \mid \mathbf{y}_i).
\end{equation}

Consensus Monte Carlo allows us to obtain the global posterior distribution by combining the  samples via weighted averages. For the vector of hyperparameters $\boldsymbol\theta$, we generate  a series of $S$ draws as follows:
\begin{equation}
    \boldsymbol\theta_s = \left(\sum_{i=1}^n w_i\right)^{-1} \sum_{i=1}^n w_{i} \boldsymbol\theta_{i,s},
\end{equation}
where the weights $w_i = \tilde{\pi}(\mathbf{y}_i)$ correspond to the marginal likelihood of each  partition. From these samples, we can derive the empirical marginal posterior distribution for the  $j$-th hyperparameter, $\theta_j$, using the set of draws $\{\theta_{j1}, \dots, \theta_{jS}\}$.

\section{Recursive inference}
\label{sec:recursive_inference}

The recursive inference approach shares many conceptual steps with the previously described distributed framework. Building on the results already established, we now summarize the relevant sections from the distributed setting, adapting them to the recursive context. Our focus will be on the specific procedures and features that distinguish recursive inference, while referencing shared components only as needed. This allows us to concentrate more thoroughly on the unique aspects of recursive inference within the INLA methodology.

Therefore, as in the distributed framework, we will examine how to obtain the posterior distributions in the recursive approach: (i) the conditional posterior distribution of the latent field, $\pi(\mathbf{x} \mid \boldsymbol\theta, \mathbf{y})$; (ii) the joint marginal posterior distribution of the hyperparameters, $\pi(\boldsymbol\theta \mid \mathbf{y})$; (iii) the marginal posterior distribution of each latent field node, $\pi(x_i \mid \mathbf{y})$; and (iv) the marginal posterior distribution of each hyperparameter, $\pi(\theta_j \mid \mathbf{y})$.

\subsection{Conditional posterior distribution of the latent field}

The conditional posterior distribution of the latent field is a fundamental component for implementing recursive inference. As detailed in Section~\ref{sec:cond_post_distributed}, we distinguish between two scenarios based on the likelihood type. In the Gaussian case, an exact identity exists between the standard and recursive inferential approaches. Conversely, for non-Gaussian likelihoods, the recursive formulation yields an approximation analogous to the one used in the distributed inference framework.

In the Gaussian likelihood case, we can proceed analogously to Eq.~(\ref{eq:conditional_distributed_extended}), restructuring the expression to show that the prior of the latent field at the $i$-th step of the recursive inference is updated using the conditional posterior of the latent field from step $(i-1)$:
\begin{equation}
    \resizebox{.9\textwidth}{!}{
    $
    \begin{array}{rcl}
        \log\pi(\mathbf{x} \mid \mathbf{y}_1, \dots, \mathbf{y}_i, \boldsymbol\theta) & = & \displaystyle C(\boldsymbol\theta, \mathbf{y}) + \sum_{j=1}^i\log\pi(\mathbf{y}_j \mid \mathbf{A}_j\mathbf{x}, \boldsymbol\theta) + \log\pi(\mathbf{x} \mid \boldsymbol\theta_2)\\
        & = & \displaystyle C^{\prime}(\boldsymbol\theta, \mathbf{y}) + \log\pi(\mathbf{y}_i \mid \mathbf{A}_i\mathbf{x}, \boldsymbol\theta) +\log\pi(\mathbf{x} \mid \mathbf{y}_1, \dots, \mathbf{y}_{i-1}, \boldsymbol\theta)
    \end{array}
    $
    }
    \label{eq:cond_post_Gauss_like_recursive_1}
\end{equation}
where $C(\boldsymbol\theta, \mathbf{y})$ and $C^{\prime}(\boldsymbol\theta, \mathbf{y})$ are normalizing constants ensuring equivalence. 

Therefore, the explicit expression that establishes the equivalence between the recursive approach in Eq.~(\ref{eq:cond_post_Gauss_like_recursive_1}) and the standard inferential approach for the full dataset $\mathbf{y} = \{\mathbf{y}_1, \dots, \mathbf{y}_n\}$, in the case of Gaussian likelihoods, is:
\begin{equation}
    \def\arraystretch{1.5}
    \resizebox{.9\textwidth}{!}{
    $
    \begin{array}{rcl}
        \log\pi(\mathbf{x} \mid \mathbf{y}, \boldsymbol\theta) & = & \displaystyle C(\boldsymbol\theta, \mathbf{y}) - \frac{1}{2}(\tau \cdot \mathbf{x}^{\top}\mathbf{A}^{\top}\mathbf{A}\mathbf{x} - 2\cdot \tau \cdot \mathbf{y}^{\top}\mathbf{A}\mathbf{x}) - \frac{1}{2}\mathbf{x}^{\top}\mathbf{Q}(\boldsymbol\theta_2)\mathbf{x}  \\
         & = & \displaystyle C(\boldsymbol\theta, \mathbf{y}) - \frac{1}{2}\sum_{j=1}^n(\tau \cdot \mathbf{x}^{\top}\mathbf{A}_j^{\top}\mathbf{A}_j\mathbf{x} - 2\cdot \tau \cdot \mathbf{y}_j^{\top}\mathbf{A}_j\mathbf{x}) - \frac{1}{2}\mathbf{x}^{\top}\mathbf{Q}(\boldsymbol\theta_2)\mathbf{x} \\
         & = & \displaystyle C(\boldsymbol\theta, \mathbf{y}) - \frac{1}{2}\sum_{j=i}^n(\tau \cdot \mathbf{x}^{\top}\mathbf{A}_j^{\top}\mathbf{A}_j\mathbf{x} - 2\cdot \tau \cdot \mathbf{y}_j^{\top}\mathbf{A}_j\mathbf{x}) \\ & & \displaystyle -\frac{1}{2}\left(\sum_{m=1}^{i-1}(\tau \cdot \mathbf{x}^{\top}\mathbf{A}_m^{\top}\mathbf{A}_m\mathbf{x} - 2\cdot \tau \cdot \mathbf{y}_m^{\top}\mathbf{A}_m\mathbf{x}) + \mathbf{x}^{\top}\mathbf{Q}(\boldsymbol\theta_2)\mathbf{x}\right) \\
         & = & \displaystyle C^{\prime}(\boldsymbol\theta, \mathbf{y}) + \sum_{j=i}^n\log\pi(\mathbf{y}_j \mid \mathbf{A}_j\mathbf{x}, \boldsymbol) + \log\pi(\mathbf{x} \mid \mathbf{y}_1,...,\mathbf{y}_{i-1}, \boldsymbol\theta) \\
         & = & \displaystyle C^*(\boldsymbol\theta, \mathbf{y}) + \log\pi(\mathbf{y}_n \mid \mathbf{A}_n\mathbf{x}, \boldsymbol) + \log\pi(\mathbf{x} \mid \mathbf{y}_1,...,\mathbf{y}_{n-1}, \boldsymbol\theta)
    \end{array}$}
    \label{eq:conditional_recursive_extended}
\end{equation}
where $C(\boldsymbol\theta, \mathbf{y})$, $C^{\prime}(\boldsymbol\theta, \mathbf{y})$ and $C^*(\boldsymbol\theta, \mathbf{y})$ are again constants ensuring equivalence. The conditional posterior distributions $\pi(\mathbf{x} \mid \mathbf{y}_1, \dots, \mathbf{y}_{i-1}, \boldsymbol\theta)$ and $\pi(\mathbf{x} \mid \mathbf{y}_1, \dots, \mathbf{y}_{n-1}, \boldsymbol\theta)$ serve as priors for the $i$-th and $n$-th steps, respectively, in the recursive framework. This remarks that the recursive conditional posterior is equal to the one obtained from the standard approach and that it is invariant to the ordering in which observations are introduced recursively.

In general, for non-Gaussian likelihoods, the conditional posterior at the $i$-th step of the recursive approach is approximated via a second-order Taylor expansion of the log-likelihood around the mode of the true conditional posterior $\pi(\mathbf{x} \mid \mathbf{y}_1, \dots, \mathbf{y}_i, \boldsymbol\theta)$. This Gaussian approximation---used also in the standard inferential approach based on the full dataset---computes the mode $\boldsymbol\mu(\boldsymbol\theta)$ of the conditional latent field as the minimizer of the negative log conditional posterior:
$$
\resizebox{.9\textwidth}{!}{$
\boldsymbol\mu(\boldsymbol\theta) = \underset{\mathbf{x}}{\argmin}\left\lbrace -\log\pi(\mathbf{x} \mid \mathbf{y}, \boldsymbol\theta) \right\rbrace = \underset{\mathbf{x}}{\argmin}\left\lbrace -\log \pi(\mathbf{y} \mid \mathbf{x}, \boldsymbol\theta) -\log\pi(\mathbf{x} \mid \boldsymbol\theta) \right\rbrace.
$}
$$

This is equivalent to stating that $\mathbf{x} = \boldsymbol\mu(\boldsymbol\theta)$ is the point at which the gradient of the negative log conditional posterior is equal to zero:
$$
\nabla_{\mathbf{x}}\left( -\log\pi(\mathbf{x} \mid \mathbf{y}, \boldsymbol\theta)\right)\vert_{\mathbf{x}=\boldsymbol\mu(\boldsymbol\theta)} = -\nabla_{\mathbf{x}} \log \pi(\mathbf{y} \mid \mathbf{x}, \boldsymbol\theta)\vert_{\mathbf{x}=\boldsymbol\mu(\boldsymbol\theta)} - \nabla_{\mathbf{x}} \log\pi(\mathbf{x} \mid \boldsymbol\theta)\vert_{\mathbf{x}=\boldsymbol\mu(\boldsymbol\theta)} =\mathbf{0},
$$
so that, for a prior distribution $\pi(\mathbf{x} \mid \boldsymbol\theta_2) = \text{GMRF}(\mathbf{0}, \mathbf{Q}(\boldsymbol\theta_2))$, the mode satisfies:
\begin{equation}
-\nabla_{\mathbf{x}}\log\pi(\mathbf{y} \mid \mathbf{x}, \boldsymbol\theta)\vert_{\mathbf{x}=\boldsymbol\mu(\boldsymbol\theta)} + (\mathbf{Q}(\boldsymbol\theta_2)\mathbf{x})\vert_{\mathbf{x}=\boldsymbol\mu(\boldsymbol\theta)} = \mathbf{0}.
\end{equation}

This equation does not admit a closed-form solution, except for the case of Gaussian likelihoods, and must be solved using numerical optimization techniques. These may include Newton-type methods---which require explicit evaluation of the Hessian at each iteration (and thus benefit from known analytic expressions of it to accelerate convergence)---or quasi-Newton methods such as \textsc{L-BFGS}, which avoid the explicit computation of the Hessian. Other numerical optimization techniques can also be used. In particular, the \texttt{R-INLA} software employs Newton-Raphson methods that are enhanced with modifications to both the gradient and Hessian to improve convergence and exploration. These enhancements are implemented through the \textit{Smart Gradient} and \textit{Smart Hessian} strategies proposed in \cite{Fattah_2022_smartGradientHessian}.

The expression for the Hessian, which corresponds to the precision matrix of the Gaussian approximation, is:
\begin{equation}
    \def\arraystretch{1.5}
    \begin{array}{rcl}
        \mathbf{Q}_{\mathbf{x} \mid \mathbf{y}, \boldsymbol\theta}(\boldsymbol\theta) & = & \displaystyle \nabla^2_{\mathbf{x}}(- \log\pi(\mathbf{x} \mid \mathbf{y}, \boldsymbol\theta)) \\
         & = & \displaystyle \mathbf{Q}_{\mathbf{x}\mid \boldsymbol\theta_2}(\boldsymbol\theta_2) - \nabla_{\mathbf{x}}^{2}\log\pi(\mathbf{y} \mid \mathbf{x}, \boldsymbol\theta) \\
         & = & \displaystyle \mathbf{Q}_{\mathbf{x}\mid \boldsymbol\theta_2}(\boldsymbol\theta_2) - \mathbf{A}^{\top} \nabla_{\boldsymbol\eta}^{2}\log\pi(\mathbf{y} \mid \boldsymbol\eta, \boldsymbol\theta) \mathbf{A}
    \end{array}
\end{equation}
where $\boldsymbol\eta = \mathbf{A}\mathbf{x}$ is the linear predictor. This formulation simplifies the expression for the curvature of the likelihood since, under conditional independence, the likelihood factorizes as $\pi(\mathbf{y} \mid \boldsymbol\eta, \boldsymbol\theta) = \prod_{i=1}^n \pi(\mathbf{y}_i \mid \eta_i, \boldsymbol\theta)$, leading to a diagonal structure for the matrix $\nabla_\eta^{2}\log\pi(\mathbf{y} \mid \boldsymbol\eta, \boldsymbol\theta)$, denoted by $\mathbf{D}$. This is particularly relevant when evaluating the conditional posterior at the mode.

In the recursive framework, these steps are carried out sequentially over the data partitions $\mathbf{y} = {\mathbf{y}_1, \dots, \mathbf{y}_n}$. At each step, the conditional prior for $\mathbf{x}$, used in the analysis of $\mathbf{y}_i$, is given by the conditional posterior from the previous step, $\pi(\mathbf{x} \mid \mathbf{y}_1, \dots, \mathbf{y}_{i-1}, \boldsymbol\theta)$. Accordingly, the conditional posterior at step $i$ is approximated as:
\begin{equation}
    \def\arraystretch{1.5}
    \resizebox{.9\textwidth}{!}{
    $
    \begin{array}{rcl}
         \log\pi_G(\mathbf{x} \mid \mathbf{y}_1, \dots, \mathbf{y}_i, \boldsymbol\theta) & = & \displaystyle\log\mathbf{C}(\boldsymbol\theta, \mathbf{y}_1, \dots, \mathbf{y}_i) + \\
         & & \displaystyle \nabla_{\mathbf{x}}\log\pi(\mathbf{x \mid \mathbf{y}_1, \dots, \mathbf{y}_i, \boldsymbol\theta}) \vert_{\mathbf{x}=\mathbf{\boldsymbol\mu_i(\boldsymbol\theta)}} \mathbf{x} + \\
         & & \displaystyle \frac{1}{2}\mathbf{x}^{\top}\nabla^2_x\log\pi(\mathbf{x \mid \mathbf{y}_1, \dots, \mathbf{y}_i, \boldsymbol\theta}) \vert_{\mathbf{x}=\mathbf{\boldsymbol\mu_i(\boldsymbol\theta)}} \mathbf{x} \\
         & = & \log\mathbf{C}(\boldsymbol\theta, \mathbf{y}_1, \dots, \mathbf{y}_i) -\frac{1}{2}(\mathbf{x} - \boldsymbol\mu_i(\boldsymbol\theta))^{\top} \mathbf{Q}_i(\boldsymbol\theta)(\mathbf{x} - \boldsymbol\mu_i(\boldsymbol\theta)).
    \end{array}$}
    \label{eq:cond_post_recursive}
\end{equation}

In this equation, the precision matrix $\mathbf{Q}_i(\boldsymbol\theta)$ is updated recursively as:
\begin{equation}
    \mathbf{Q}_i(\boldsymbol\theta) = \mathbf{Q}_{i-1}(\boldsymbol\theta) + \mathbf{A}_i^{\top}\nabla_{\boldsymbol\eta}^2\log\pi(\mathbf{y}_i \mid \boldsymbol\eta_i, \boldsymbol\theta)\vert_{\mathbf{x}=\boldsymbol\mu_i(\boldsymbol\theta)}\mathbf{A}_i = \mathbf{Q}_{i-1}(\boldsymbol\theta) + \mathbf{A}_i^{\top}\mathbf{D}_i(\boldsymbol\theta, \mathbf{y}_i)\mathbf{A}_i
    \label{eq:prec_cond_post_rec}
\end{equation}
which leads to the following explicit recurrence relation:
\begin{equation}
    \mathbf{Q}_i(\boldsymbol\theta) = \mathbf{Q}(\boldsymbol\theta_2) + \sum_{j=1}^{i}\mathbf{A}_j^{\top}\mathbf{D}_j(\boldsymbol\theta, \mathbf{y}_j)\mathbf{A}_j.
\end{equation}

The mean $\boldsymbol\mu_i(\boldsymbol\theta)$, corresponding to the mode of the true conditional posterior, is also computed recursively:
\begin{equation}
    \boldsymbol\mu_i(\boldsymbol\theta) = \underset{\mathbf{x}}{\argmin}\left\lbrace -\log \pi(\mathbf{y}_i \mid \mathbf{x}, \boldsymbol\theta) -\log\pi_G(\mathbf{x} \mid \mathbf{y}_1, \dots, \mathbf{y}_{i-1}, \boldsymbol\theta) \right\rbrace.
\end{equation}
which is equivalent to solving:
\begin{equation}
    -\nabla_{\mathbf{x}}\log\pi(\mathbf{y}_i \mid \mathbf{x}, \boldsymbol\theta) + \mathbf{Q}_{i-1}(\boldsymbol\theta) (\mathbf{x} - \boldsymbol\mu_{i-1}(\boldsymbol\theta)) = \mathbf{0}.
    \label{eq:mean_cond_post_rec}
\end{equation}

Alternatively, by defining the change of variable $\mathbf{x} = \mathbf{z} + \boldsymbol\mu_{i-1}(\boldsymbol\theta)$, we obtain:
$$
-\nabla_{\mathbf{z}}\log\pi(\mathbf{y}_i \mid \mathbf{z}, \boldsymbol\theta) + \mathbf{Q}_{i-1}(\boldsymbol\theta) \mathbf{z} = \mathbf{0}.
$$

These expressions highlight that the recursive approach is, in general, not invariant under reordering of the data partitions. This is because the conditional posterior distribution computed for non-Gaussian likelihoods is approximated by a Gaussian approximation of the true conditional posterior. Although the mode of the Gaussian approximation coincides with that of the true posterior, the geometric information of the latter is not fully captured by the approximation. As a result, when computing the modal configuration in the recursive framework using Eq.~(\ref{eq:mean_cond_post_rec}), the update does not preserve the same information and is not invariant, since the operation is not commutative. See the \emph{Supplementary Material} for a more detailed discussion on the non-invariance for non-Gaussian likelihoods.

\subsection{Joint marginal posterior distribution of the hyperparameters}

The next step is to compute the joint marginal posterior distribution of the hyperparameters, for which we follow the standard procedure by evaluating the functional expression for $\pi(\boldsymbol\theta \mid \mathbf{y})$ given in Eq.~(\ref{eq:marg_hyperpar_Std_INLA}).

In the recursive approach, as in the distributed one, there are two alternatives for approximating the joint marginal posterior. One option is to directly use Eq.(\ref{eq:marg_hyperpar_Std_INLA}), computing:
\begin{equation}
    \tilde{\pi}(\boldsymbol\theta \mid \mathbf{y}) \propto \frac{\pi(\mathbf{y}_1, \dots, \mathbf{y}_n \mid \mathbf{x}, \boldsymbol\theta)\pi(\mathbf{x} \mid \boldsymbol\theta)\pi(\boldsymbol\theta)}{\pi_G(\mathbf{x} \mid \mathbf{y}_1, \dots, \mathbf{y}_n, \boldsymbol\theta)} 
    \label{eq:hyperpar_post_recursive_1}
\end{equation}
where $\pi_G(\mathbf{x} \mid \mathbf{y}_1, \dots, \mathbf{y}_n, \boldsymbol\theta)$ denotes the conditional posterior distribution of the latent field computed through the recursive inference approach, as described in Eq.~(\ref{eq:cond_post_recursive}). In this case, model constraints Eq.~(\ref{eq:conditioning_kriging}) must be applied to the Gaussian approximation $\pi_G(\mathbf{x} \mid \mathbf{y}_1, \dots, \mathbf{y}_n, \boldsymbol\theta)$.

An alternative approach for approximating the joint marginal posterior of the hyperparameters leverages the information already computed and defines a recursive procedure to approximate $\pi(\boldsymbol\theta \mid \mathbf{y})$. In this case, the recursive expression is:
\begin{equation}
    \tilde{\pi}_r(\boldsymbol\theta \mid \mathbf{y}) \propto \left[\prod_{i=1}^n\frac{\pi(\mathbf{y}_i \mid \mathbf{x}, \boldsymbol\theta)\pi_G(\mathbf{x} \mid \mathbf{y}_1, \dots, \mathbf{y}_{i-1}, \boldsymbol\theta)}{\pi_G(\mathbf{x} \mid \mathbf{y}_1, \dots, \mathbf{y}_{i}, \boldsymbol\theta)}\right]_{\mathbf{x}=\boldsymbol\mu_{i}(\boldsymbol\theta)} \pi(\boldsymbol\theta)
    \label{eq:hyperpar_post_recursive_2}
\end{equation}
where, for each conditional posterior distribution $\pi_G(\mathbf{x} \mid \mathbf{y}_1, \dots, \mathbf{y}_{i}, \boldsymbol\theta)$ appearing in the numerator and denominator, the constraints defined by Eq.~(\ref{eq:conditioning_kriging}) must also be applied.

\subsection{Marginal posterior distribution of the latent field}

The marginal posterior distribution of the latent field nodes is one of the key components of the INLA framework. To compute it, it is necessary to build an integration scheme for evaluating the marginal distributions using Eq.~(\ref{eq:marginal_posterior_latentfield}). A required step for constructing such a scheme is to find the mode of the marginal posterior distribution of the hyperparameters, typically through optimization techniques based on finite differences, since the distribution is not analytically differentiable. Once this mode is identified, one can carry out an exploration strategy (e.g., a grid or central composite design, CCD) of the hyperparameter space to obtain both the marginal distributions of the latent field nodes and the marginal posteriors of the hyperparameters.

In the corresponding section of the distributed framework, we discussed several ways to construct these integration schemes by leveraging the structure of the distributed inference itself. However, in the recursive setting, building an integration scheme directly from the recursive approach is unfeasible if an accurate localization of the mode is desired. While one could, in principle, use the mode estimated from the first data subset, in practice this is inadequate, as the mode will likely shift as new data are incorporated. Thus, the only viable option for a recursive design is to use the distributed strategy in a sequential fashion, i.e., applying it as new data subsets become available.

The construction of the integration scheme---particularly the CCD---can be based on the same strategy discussed for the distributed framework, using Eqs.~(\ref{eq:dist_hyper_to_CCD_1}) through (\ref{eq:dist_hyper_to_CCD_4}). With these, for each dataset $\mathbf{y}_i$ corresponding to the $i$-th step of the recursive procedure, we can build a Gaussian approximation centered at the modal configuration of the hyperparameters for that specific dataset. That is, for each $\mathbf{y}_i$ we construct $\pi_G(\boldsymbol\theta \mid \mathbf{y}_i)$, and combine the different approximations to obtain a modal approximation for the hyperparameters associated with the combined datasets $\{\mathbf{y}_1, \dots, \mathbf{y}_i\}$ as $\pi_G(\boldsymbol\theta \mid \mathbf{y}_1, \dots, \mathbf{y}_i)\approx \prod_{j=1}^{i}\pi_G(\boldsymbol\theta \mid \mathbf{y}_j)$, from which we can then build the appropriate integration scheme for that step of the recursive procedure.

Once the integration scheme is defined---for instance, a CCD scheme without loss of generality---with support points $\{\boldsymbol\theta^{k}\}_{k=1}^K$, we can use either Eq.~(\ref{eq:hyperpar_post_recursive_1}), using the conditional posterior of the latent field obtained via the recursive approach, or Eq.~(\ref{eq:hyperpar_post_recursive_2}) to compute the marginal posterior distribution of the hyperparameters. After evaluating this posterior and having computed the conditional posterior of the latent field (also through the recursive method) at each support point, we can finally compute the marginal posterior distribution for each latent field node:
\begin{equation}
    \tilde{\pi}_r(x_j \mid \mathbf{y}_1, \dots, \mathbf{y}_i) = \sum_{k=1}^K \tilde{\pi}_G(x_j \mid \mathbf{y}_1, \dots, \mathbf{y}_i, \boldsymbol\theta^k)\tilde{\pi}_r(\boldsymbol\theta^k \mid \mathbf{y}_1, \dots, \mathbf{y}_i) \Delta_k,
    \label{eq:rec_marginal_nodes}
\end{equation}
where $\tilde{\pi}_r(\boldsymbol\theta^k \mid \mathbf{y}_1, \dots, \mathbf{y}_i)$ is the marginal posterior of the hyperparameters at support point $\boldsymbol\theta^k$, and $\Delta_k$ are the integration weights. The conditional posterior of the latent field and its marginal at this stage incorporate both the constraints and the low-rank correction.

It is important to note that, due to changes in the integration support points, the information computed in the $(i-1)$-th step of the recursive approach cannot, in general, be reused in step $i$. This is because both the conditional posterior of the latent field, $\pi_G(\mathbf{x} \mid \mathbf{y}_1, \dots, \mathbf{y}_{i-1}, \boldsymbol\theta^k)$, and the marginal posterior of the hyperparameters, $\tilde{\pi}_r(\boldsymbol\theta^k \mid \mathbf{y}_1, \dots, \mathbf{y}_{i-1})$, depend on the specific integration support points $\{\boldsymbol\theta^k\}_{k=1}^K$, which may change as $\pi_G(\boldsymbol\theta \mid \mathbf{y}_1, \dots, \mathbf{y}_i)$ is recomputed as a product of the Gaussians $\pi_G(\boldsymbol\theta \mid \mathbf{y}_j)$.

However, what can be reused throughout the recursive process is the Gaussian approximation of the hyperparameter mode for each individual dataset, since this approximation is computed independently for each $\mathbf{y}_i$. Moreover, if the Gaussian approximation of the hyperparameter mode at step $i$ is sufficiently close to that from step $(i-1)$, then the integration scheme, and all associated quantities (the conditional posterior of the latent field and the marginal posterior of the hyperparameters), can also be reused from the previous step.

\subsection{Marginal posterior distribution of the hyperparameters}

The marginal posterior distributions of each hyperparameter, $\tilde{\pi}(\theta_i \mid \mathbf{y})$, can be computed using one of four strategies described for the distributed inference framework. These strategies can be grouped into two categories: those that rely on interpolation---namely, (i) interpolation and (ii) asymmetric Gaussian interpolation---and those that do not, such as (iii) Laplace approximation for each hyperparameter $\theta_j$, (iv) the numerical integration-free algorithm, and (v) consensus Monte Carlo, which leverages the application of the numerical integration-free algorithm used in each partition to approximate the joint posterior marginal distribution of the hyperparameters through an asymmetric Gaussian approximation.

Among these, the most practical for recursive implementation are the (i) simple interpolation method, (iv) numerical integration-free algorithm, and (v) consensus Monte Carlo. The former is especially convenient when the posterior exploration is restricted to a low-dimensional subspace $\boldsymbol\Theta_0 \subset \boldsymbol\Theta$, which can be covered with a fixed, well-chosen set of support points $\{\boldsymbol\theta^k\}_{k=1}^K$. On the other hand, the numerical integration-free algorithm and the consensus Monte Carlo offer the lowest computational cost among the alternatives, requiring only the Gaussian approximation at the mode and three evaluations along each side of every hyperparameter dimension, and drawing samples from the approximated joint posterior distributed for each partition in the case of the consensus Monte Carlo. This implies the use of the expression for the posterior density of the hyperparameters either in the recursive framework or in the distributed framework.

\section{Partitioning of the latent field}

In the previous two sections, we presented in detail the distributed and recursive inference approaches, leveraging the INLA methodology. Both frameworks assume that the data are divided into subsets, $\mathcal{P}(\mathbf{y}) = \{\mathbf{y}_1, \dots, \mathbf{y}_n\}$. However, we have not yet specified how to partition the data when the entire dataset---or a large portion of it---is available from the outset. Therefore, this section is dedicated to detailing latent field partitioning within both distributed and recursive frameworks.

In this context, we are interested in reducing the computational burden by partitioning not only the data but also the latent field or model structure itself. To this end, the model structure can be partitioned according to the natural partitioning of the data, $\mathcal{P}(\mathbf{y}) \rightarrow \mathcal{P}(\mathbf{x})$, provided that the data structure permits it. Alternatively, one can first partition the latent field and subsequently derive the data partitions from it, $\mathcal{P}(\mathbf{x}) \rightarrow \mathcal{P}(\mathbf{y})$. We assume for simplicity that this partitioning is performed exclusively based on a specific component of the model, although it is possible to use two or more components to perform the partition, e.g. a spatio-temporal partition of the model. That is, consider a linear predictor defined as $\boldsymbol\eta = \mathbf{A}\mathbf{x}$, where $\mathbf{A} = (\mathbf{A}_1, \dots, \mathbf{A}_J)$ and $\mathbf{x}^{\top} = (\mathbf{x}_1^\top, \dots, \mathbf{x}_J^\top)$, with each matrix $\mathbf{A}_j$ representing the projection matrix associated with the $j$-th component $\mathbf{x}_j$ of the latent field (e.g., spatial, temporal, spatio-temporal, or other random effects). The partitioning of the entire latent field $\mathbf{x}$ is then carried out by focusing solely on the structure of a single chosen component $\mathbf{x}_j$.

Therefore, the purpose of partitioning extends beyond merely subdividing observations into manageable subsets for distributed or recursive inference; it also involves a structured decomposition of the latent field itself. This aspect is central to the design of the partitioning scheme and directly relates to Eqs.~\eqref{eq:conditional_distributed_extended} and \eqref{eq:conditional_recursive_extended}. Two main schemes are adopted: (i) block-independent partitions, a naive approach that neglects correlations between adjacent nodes across partitions, or (ii) block-correlated partition, an extended formulation that incorporates such dependencies within each partition. The latter ensures that the global prior can be coherently reconstructed as the individual partitions are integrated in a distributed or recursive setting:
\begin{equation}
    \pi(\mathbf{x} \mid \boldsymbol\theta) = \prod_{i=1}^n \pi(\mathbf{x}_i \mid \boldsymbol\theta),
\end{equation}
where the latent field $\mathbf{x} \mid \boldsymbol\theta$ is a GMRF with mean vector $\boldsymbol\mu$ (typically zero) and precision matrix $\mathbf{Q}$. Similarly, each partition-specific prior $\pi(\mathbf{x}_i \mid \boldsymbol\theta)$ is a GMRF with mean $\boldsymbol\mu_i$ and precision matrix $\mathbf{Q}_i$.

Both schemes involve distinguishing components affected by the partitioning from those that are not, $\mathbf{x} = (\mathbf{x}_{-\mathcal{P}}, \mathbf{x}_{\mathcal{P}})$. Under both strategies, for the components used as the criterion to construct the partition, $\mathbf{x}_{\mathcal{P}}$, the precision matrix and mean vector associated with their prior distribution undergo a modification. Conversely, those components not used to define the partitioning criterion, $\mathbf{x}_{-\mathcal{P}}$, will have their prior distribution remain unaffected. The distinction can be stated by decomposing the latent field vector as $\mathbf{x} = (\mathbf{x}_{-\mathcal{P}}, \mathbf{x}_{\mathcal{P}})$.

\begin{figure}[h!]
    \centering
    \begin{subfigure}[b]{\textwidth}
        \centering
        \includegraphics[width = 0.32\linewidth]{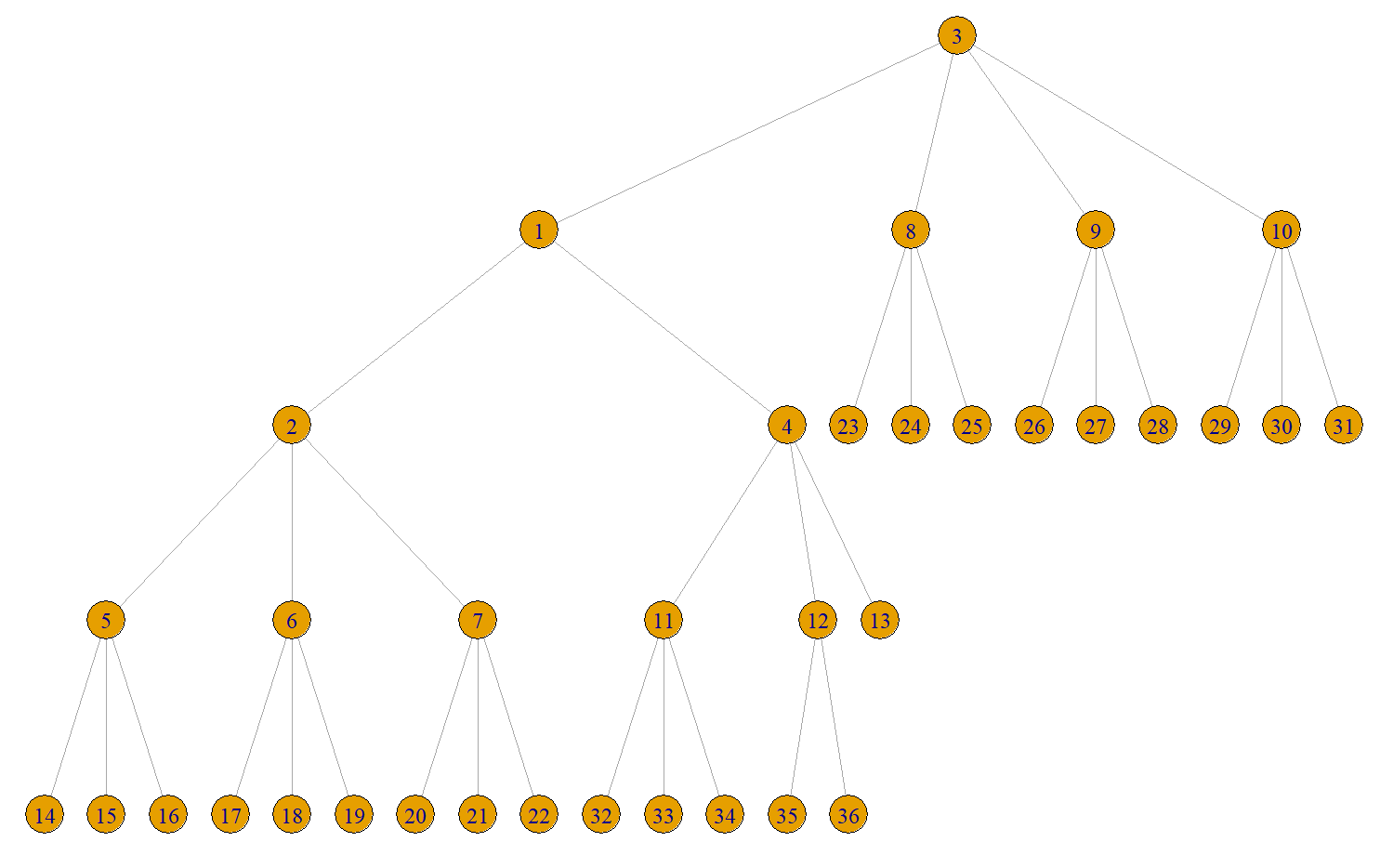}
        \hfill
        \includegraphics[width = 0.32\linewidth]{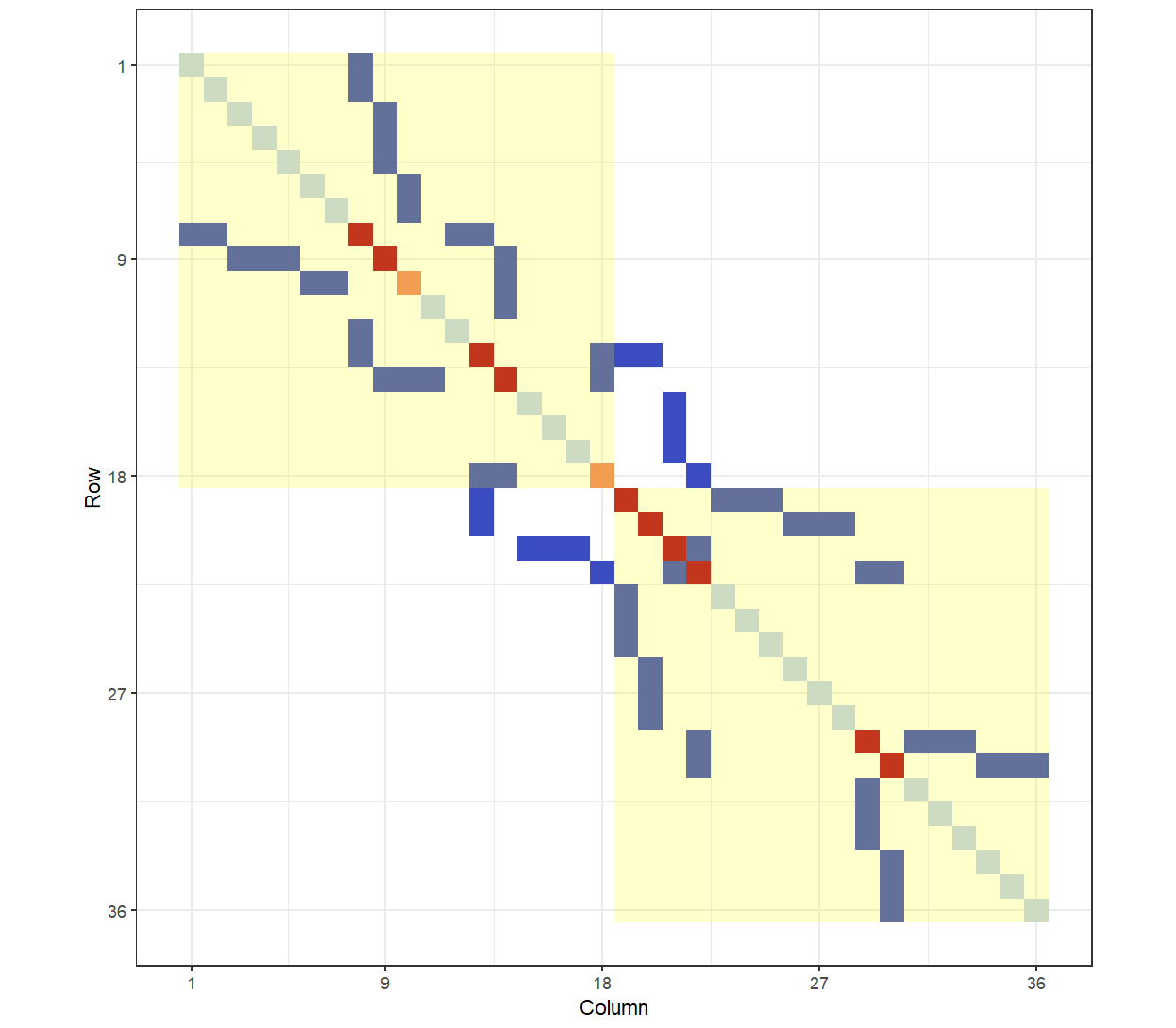}
        \hfill
        \includegraphics[width = 0.32\linewidth]{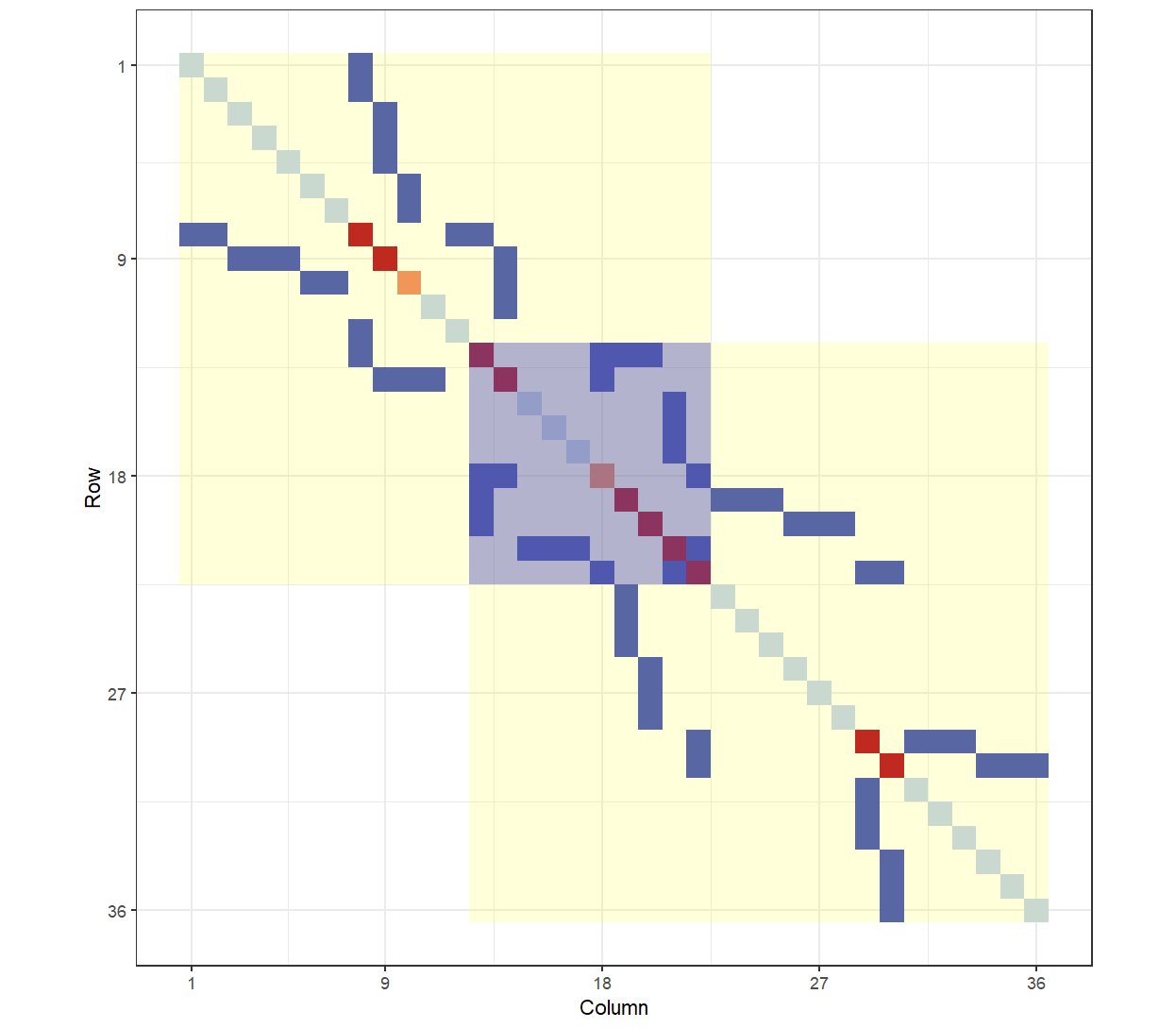}
        \caption{Dendogram example, with the graph (left), Laplacian matriz with partition highlited in yellow (middle) and Laplacian matrix with the extended partition with overlapping highlited in blue (right).}
    \end{subfigure}
    \vskip\baselineskip
    \centering
    \begin{subfigure}[b]{\textwidth}
        \centering
        \includegraphics[width = 0.32\linewidth]{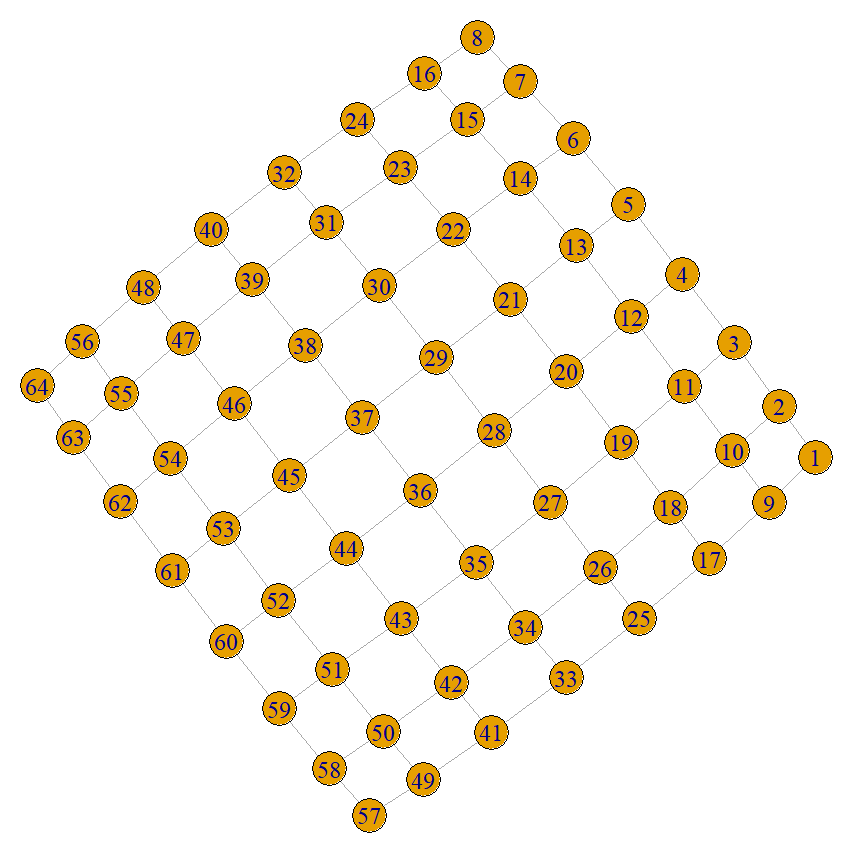}
        \hfill
        \includegraphics[width = 0.32\linewidth]{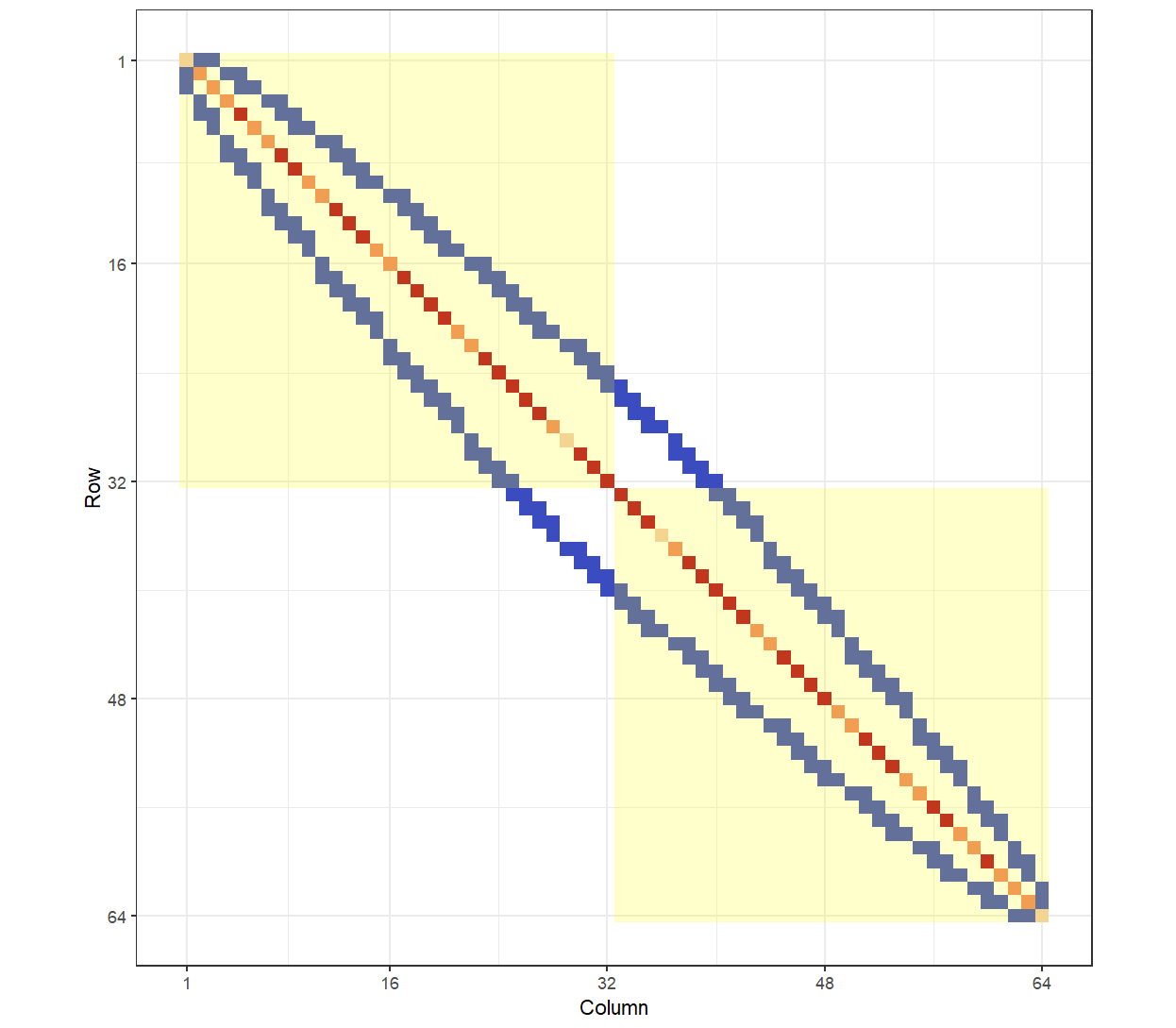}
        \hfill
        \includegraphics[width = 0.32\linewidth]{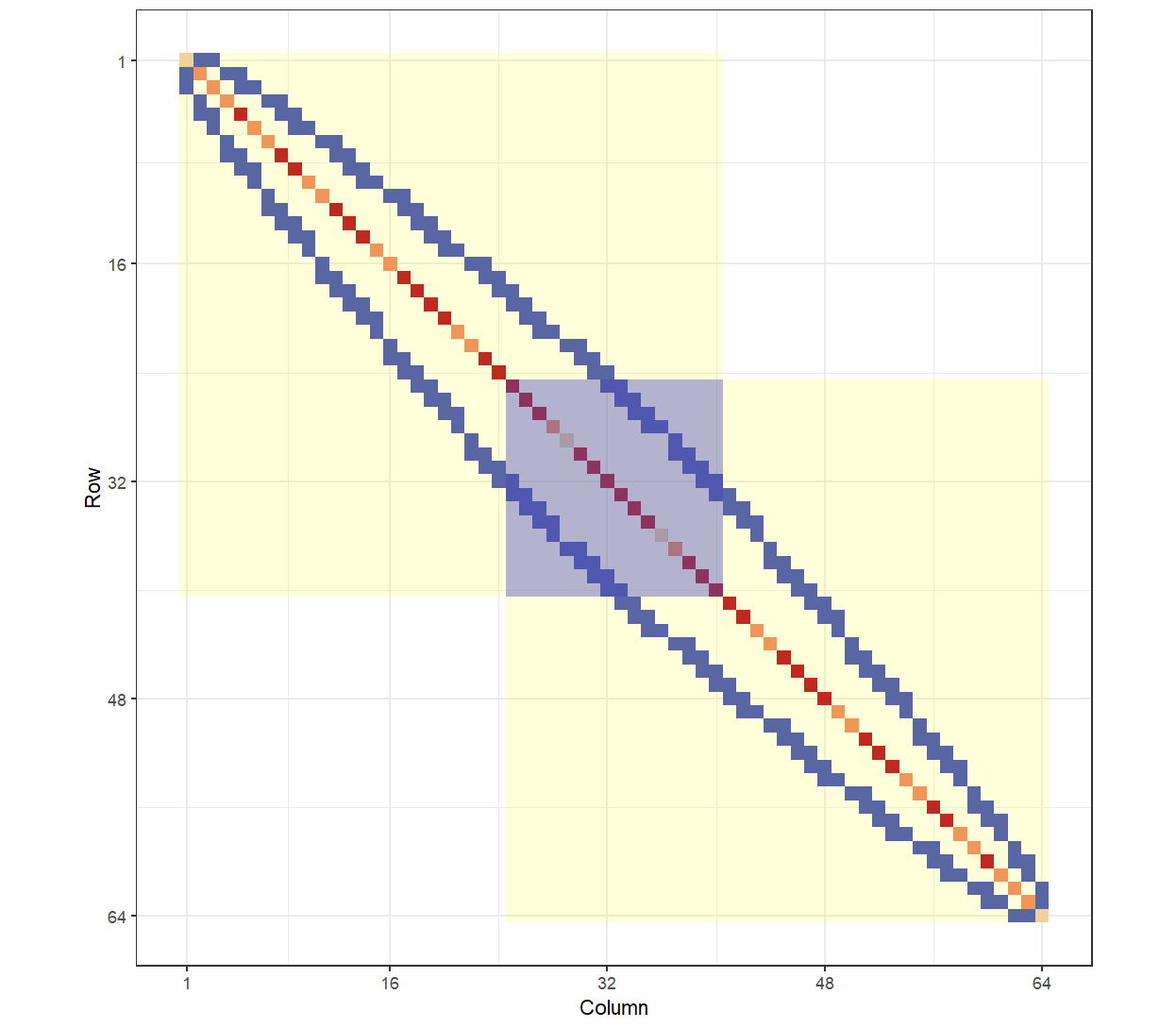}
        \caption{Grid example, with the graph (left), Laplacian matriz with partition highlited in yellow (middle) and Laplacian matrix with the extended partition with overlapping highlited in blue (right).}
    \end{subfigure}
    \caption{Example with two graphs—a dendrogram and a grid—with the diagonal blocks in the Laplacian matrix and the extended blocks capturing the interactions between the original blocks, highlighting the common elements.}
    \label{fig:example_partitioning_intro}
\end{figure}

In Figure~\ref{fig:example_partitioning_intro} we have two examples for both splitting approaches. In the left we have the graph related to the component, in the middle we have a partition from the block-independeent scheme, and in the right we have a partition arising from the block-correlated scheme. We can clearly see that the block-independent loses some correlation elements from the precision matrix, which implies that we modify the underlying conditional independence propoerties defining the component to reduce the computational burden, meanwhile in the block-correlated partition we seek to keep such correlations, being able to reconstruct the overall precision matrix, therefore without altering the conditional properties of the component used to create the partition.

\subsection{Block-independent partitions}

The block-independent partition scheme offers a straightforward conceptual framework for reducing the structural complexity of a GMRF. By explicitly severing spatial, temporal, or general graph connections that cross partition boundaries, the original globally connected latent graph is decomposed into a set of mutually independent subgraphs. This independence assumption simplifies both the topological structure of the model and the downstream computational tasks.

When the primary computational bottleneck of the inference process arises from high-dimensional latent field factorizations---rather than data likelihood evaluations alone---partitioning the latent structure directly alleviates this burden. Specifically, computing the mode of the joint marginal distribution of the hyperparameters within the INLA methodology requires repeatedly evaluating the log-determinant of high-dimensional precision matrices (e.g., the prior precision matrix $\mathbf{Q}$ and the conditional posterior precision matrix $\mathbf{Q}_{\mathbf{x} \mid \boldsymbol\theta}$), an operation whose computational cost scales unfavorably with the dimension and fill-in of the graph. By downgrading the effective size of the target latent component, the block-independent formulation removes all cross-boundary couplings between adjacent partitions, enforcing a strict block-diagonal structure on the target matrix: $\mathbf{Q} \approx \mathbf{D} = \text{block-diag}(\mathbf{Q}^{(0)}_1, \dots, \mathbf{Q}^{(0)}_n)$. Consequently, the global determinant simplifies to the product of smaller partition determinants:
\begin{equation}
    |\mathbf{Q}| \approx |\mathbf{D}| = \prod_{i=1}^n |\mathbf{Q}^{(0)}_i|,
\end{equation}
which drastically reduces factorization time, memory consumption, and overall computational overhead.

This intuitive structural simplification has a formal mathematical connection to matrix perturbation analysis and random matrix theory \citep{Zhang_2005_SchurComplement, Potters_2021_RandomMatrixTheory}. Consider decomposing the full precision matrix as $\mathbf{Q} = \mathbf{D} + \mathbf{R}$, where $\mathbf{D} = \text{block-diag}(\mathbf{Q}^{(0)}_1, \dots, \mathbf{Q}^{(0)}_n)$ contains the intra-block structures and $\mathbf{R}$ gathers the off-diagonal block entries representing cross-partition interactions. The exact log-determinant can be factorized as:
\begin{equation}
    \log |\mathbf{Q}| = \log |\mathbf{D}(\mathbf{I} + \mathbf{D}^{-1}\mathbf{R})| = \log |\mathbf{D}| + \log |\mathbf{I} + \mathbf{D}^{-1}\mathbf{R}|.
\end{equation}
Using the matrix identity $\log |\mathbf{A}| = \text{tr}(\log \mathbf{A})$ and expanding the matrix logarithm via a Taylor series expansion around the identity matrix $\mathbf{I}$, we obtain:
\begin{equation}
    \log |\mathbf{Q}| = \log |\mathbf{D}| + \text{tr}\left( \mathbf{D}^{-1}\mathbf{R} \right) - \frac{1}{2}\text{tr}\left( (\mathbf{D}^{-1}\mathbf{R})^2 \right) + \dots
\end{equation}
Because $\mathbf{D}^{-1}$ is block-diagonal and $\mathbf{R}$ has zero blocks along its main diagonal, the product $\mathbf{D}^{-1}\mathbf{R}$ has zero diagonal blocks, rendering the first-order trace term identically zero: $\text{tr}(\mathbf{D}^{-1}\mathbf{R}) = 0$. Consequently, truncating this expansion at the leading terms---treating off-diagonal block interactions as small perturbations---yields 
$$
\log |\mathbf{Q}| \approx \log |\mathbf{D}| = \sum_{i=1}^n \log |\mathbf{Q}^{(0)}_i|.
$$ 
Remarkably, the determinant obtained from the block-independent scheme is mathematically identical to the determinant derived from a zeroth-order (or first-order truncated) Taylor expansion of the log-determinant in random matrix theory.

\subsection{Block-correlated partitions}

The block-independent partitioning yields substantial computational gains, but disregarding boundary correlations alters the underlying conditional independence properties of the latent field. To preserve the original dependence structure, the block-correlated partition scheme incorporates inter-block interactions directly into the local formulations. Under this scheme, evaluating the determinant is no longer tied to a simple Taylor expansion of a naive block-diagonal matrix, as cross-boundary dependencies are explicitly retained and reconstructed through extended local matrices rather than truncated. The approach leverages domain decomposition methods \citep{Toselli_2005_DomainDecomposition}, which provide the foundational background to provide an alternative way to define a splitting scheme and a log-determinant approximation. 

Unlike the block-independent scheme—where local matrices $\mathbf{Q}_i^{(0)}$ are formed by truncating connections between subdomains—the block-correlated scheme constructs extended local precision matrices $\mathbf{Q}_i$. Each extended matrix incorporates a boundary layer (or halo region) containing neighboring nodes from adjacent partitions. To ensure that overlapping elements across partitions do not double-count variances or covariances, shared entries in the precision graph are adjusted using a weight re-scaling. This guarantees an exact additive decomposition of the global precision matrix:
\begin{equation}
    \mathbf{Q} = \sum_{i=1}^n \mathbf{Q}_i.
\end{equation}

Although the sum of these extended precision matrices recovers $\mathbf{Q}$ exactly, evaluating the true global log-determinant $\log |\mathbf{Q}|$ directly can remain computationally prohibitive. To maintain efficiency, the global log-determinant is approximated as the sum of log-determinants of the extended local matrices:
\begin{equation}
    \log |\mathbf{Q}| \approx \sum_{i=1}^n \log |\mathbf{Q}_i|.
\end{equation}
This surrogate formulation is theoretically justified through a lifted space constraint perturbation framework standard in domain decomposition theory \citep{Toselli_2005_DomainDecomposition}. Specifically, the local subdomains are mapped onto an expanded (lifted) product space $\mathcal{V}_{\text{lifted}} = \bigoplus_{i=1}^n \mathbb{R}^{d_i}$ of total dimension $M = \sum_{i=1}^n d_i > N$. In this decoupled space, the block-diagonal operator $\widetilde{\mathbf{Q}} = \text{block-diag}(\mathbf{Q}_1, \dots, \mathbf{Q}_n)$ satisfies $\log |\widetilde{\mathbf{Q}}| = \sum_{i=1}^n \log |\mathbf{Q}_i|$ exactly \citep{Potters_2021_RandomMatrixTheory}. 

The true matrix $\mathbf{Q}$ on $\mathbb{R}^N$ is recovered by enforcing linear continuity constraints across overlapping interface nodes, expressed as $\mathbf{B}\widetilde{\mathbf{x}} = \mathbf{0}$, where $\mathbf{B}$ is a signed boolean constraint matrix. By the block matrix determinant identity for saddle-point systems \citep{Benzi_2005_SaddlePointProblems}, the exact log-determinant relates to the lifted log-determinant via the Schur complement of the constraint interface operator \citep{Toselli_2005_DomainDecomposition, Zhang_2005_SchurComplement}:
\begin{equation}
    \log |\mathbf{Q}| = \sum_{i=1}^n \log |\mathbf{Q}_i| - \log \left| \mathbf{B} \widetilde{\mathbf{Q}}^{-1} \mathbf{B}^\top \right| + C,
\end{equation}
where $C$ is a constant determined by space dimension projections. Consequently, adopting $\sum_{i=1}^n \log |\mathbf{Q}_i|$ as a surrogate for $\log |\mathbf{Q}|$ corresponds to a zeroth-order constraint perturbation that neglects the interface Schur complement correction $\log \left| \mathbf{B} \widetilde{\mathbf{Q}}^{-1} \mathbf{B}^\top \right|$, effectively treating inter-partition continuity constraints as weakly coupled perturbations.

In this block-correlated framework, when the global mean satisfies $\boldsymbol\mu = \mathbf{0}$, setting local partition means to $\boldsymbol\mu_i = \mathbf{0}$ ensures that the global latent field remains a zero-mean GMRF whose prior density factorizes as $\pi(\mathbf{x} \mid \boldsymbol\theta) = \prod_{i=1}^n \pi(\mathbf{x}_i \mid \boldsymbol\theta)$. In practice, setting up the precision matrix can be implemented either approximately—by enforcing a strict block-diagonal structure that disregards inter-block dependencies—or exactly, by incorporating overlapping halo zones between partitions with re-scaled matrix entries. The latter approach guarantees that cross-boundary correlations are preserved without altering the conditional properties of the latent field. Further details regarding automated graph partitioning algorithms and implementation specifications for both schemes are provided in the \emph{Supplementary Material}.

\section{Examples}

In this section, we illustrate the methodology described in Sections~\ref{sec:distributed_inference} and \ref{sec:recursive_inference}, which present the foundations of the distributed and recursive approaches. We also demonstrate the implementation of the algorithms introduced in Section~\ref{sec:algorithms_partitioning} for generating partitions based on components of the latent field. In particular, two implementation examples are presented: the first is a simple simulated spatio-temporal case with Gaussian likelihood, used to illustrate the application of the distributed and recursive inferential methods. The second is a real-data example for PM$_{2.5}$ pollutant levels in the United States, employing a more complex spatio-temporal model with space–time interaction and a non-Gaussian likelihood. 


\subsection{Distributed and recursive inference for a simple spatio-temporal model}

In this first example, we simulate a spatio-temporal dataset to illustrate, in a controlled and simple setting, the implementation of the previously described procedures for distributed and recursive inference.

The model used for data simulation is defined as follows:
\begin{equation}
\begin{array}{rcl}
     \mathbf{y} & \sim & \textbf{MVN}(\boldsymbol\mu, \tau \mathbf{I}),  \\
     \boldsymbol\mu & = & \beta_0 \mathbf{1} + \beta_{cov}\mathbf{x}_{cov} + \boldsymbol\beta_{cat} \mathbf{X}_{cat} + \mathbf{u}_s + \mathbf{u}_t, 
\end{array}
\end{equation}
where $\beta_0$ denotes the intercept, $\beta_{cov}$ is the coefficient related to the $\mathbf{x}_{cov}$ covariate values vector, $\boldsymbol\beta_{cat}$ $\mathbf{u}_s$ is a spatial component simulated using the SPDE–FEM approach, $\mathbf{}$ and $\mathbf{u}_t$ is a temporal component simulated from a first-order random walk. 

\begin{figure}
    \centering
    \includegraphics[width = 0.90\linewidth]{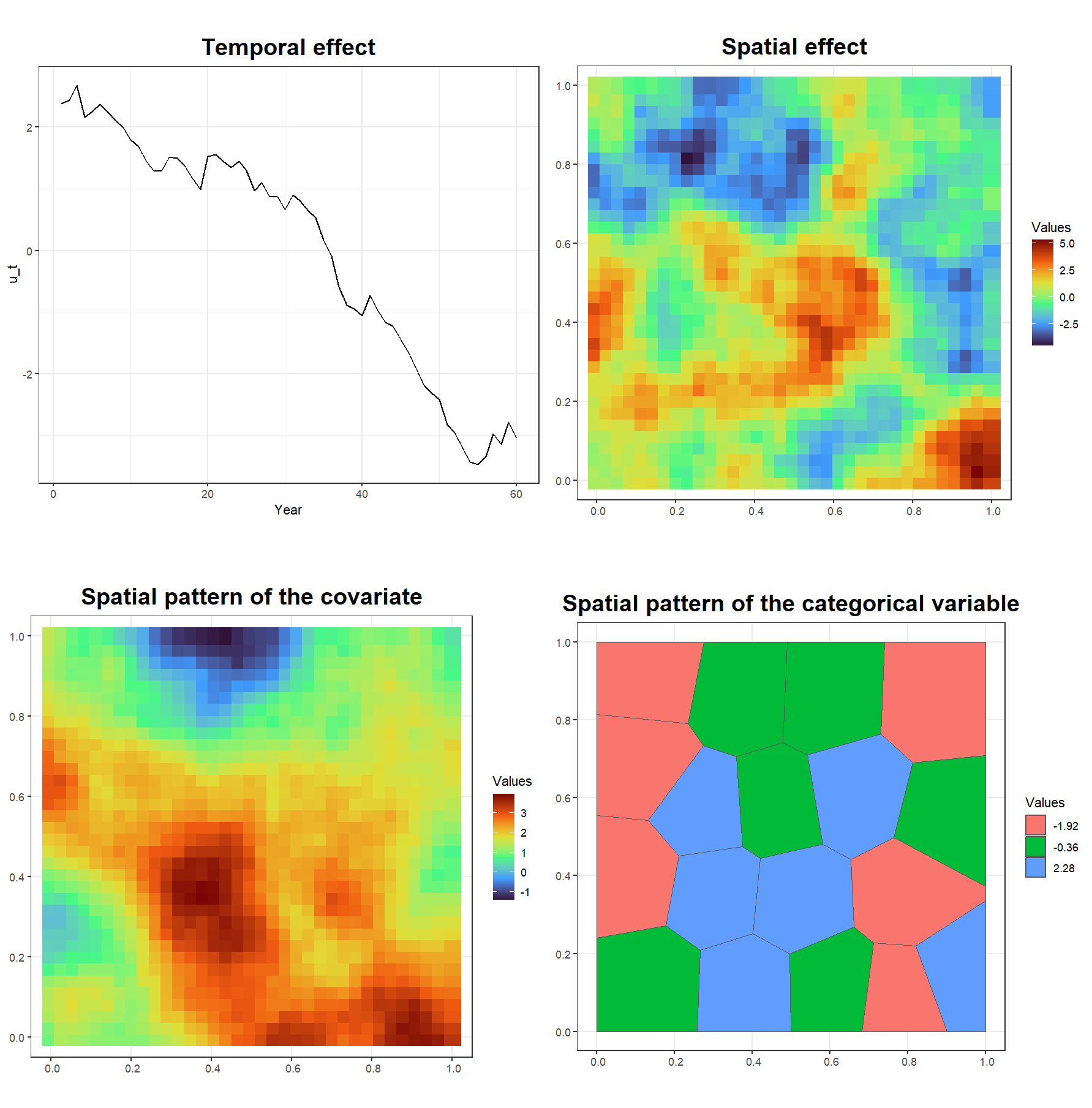}
    \caption{Plots of the temporal and spatial effects, together with the spatial patterns of the covariate and the categorical variable.}
    \label{fig:Sim_variable_plots_simple_example}
\end{figure}

In the simulation, 300 spatial locations were randomly distributed within the study region across 60 temporal nodes (years). For the simulation, the values used for the different parameters are as follow: $\beta_0= 2$, $\beta_{cov} = 1.5$ and $\boldsymbol\beta_{cat} = (-1.92, 2.28,-0.36)$ are the linear coefficients for the intercept, covariate and the levels of the categorical variable. $\tau_G = 20$ is the precision of the Gaussian likelihood, $\tau_{rw} = 20$ is the precision of the random walk, and $\sigma_{sp} = 2$ and $\rho_{sp} = 0.4$ are the marginal standard deviation and spatial range of the spatial effect (SPDE-FEM), respectively. Figure~\ref{fig:Sim_variable_plots_simple_example} shows the spatial effect, the temporal effect, and the spatial patterns of both the covariate and the categorical variable.

\begin{figure}
    \centering
    \includegraphics[width = \linewidth]{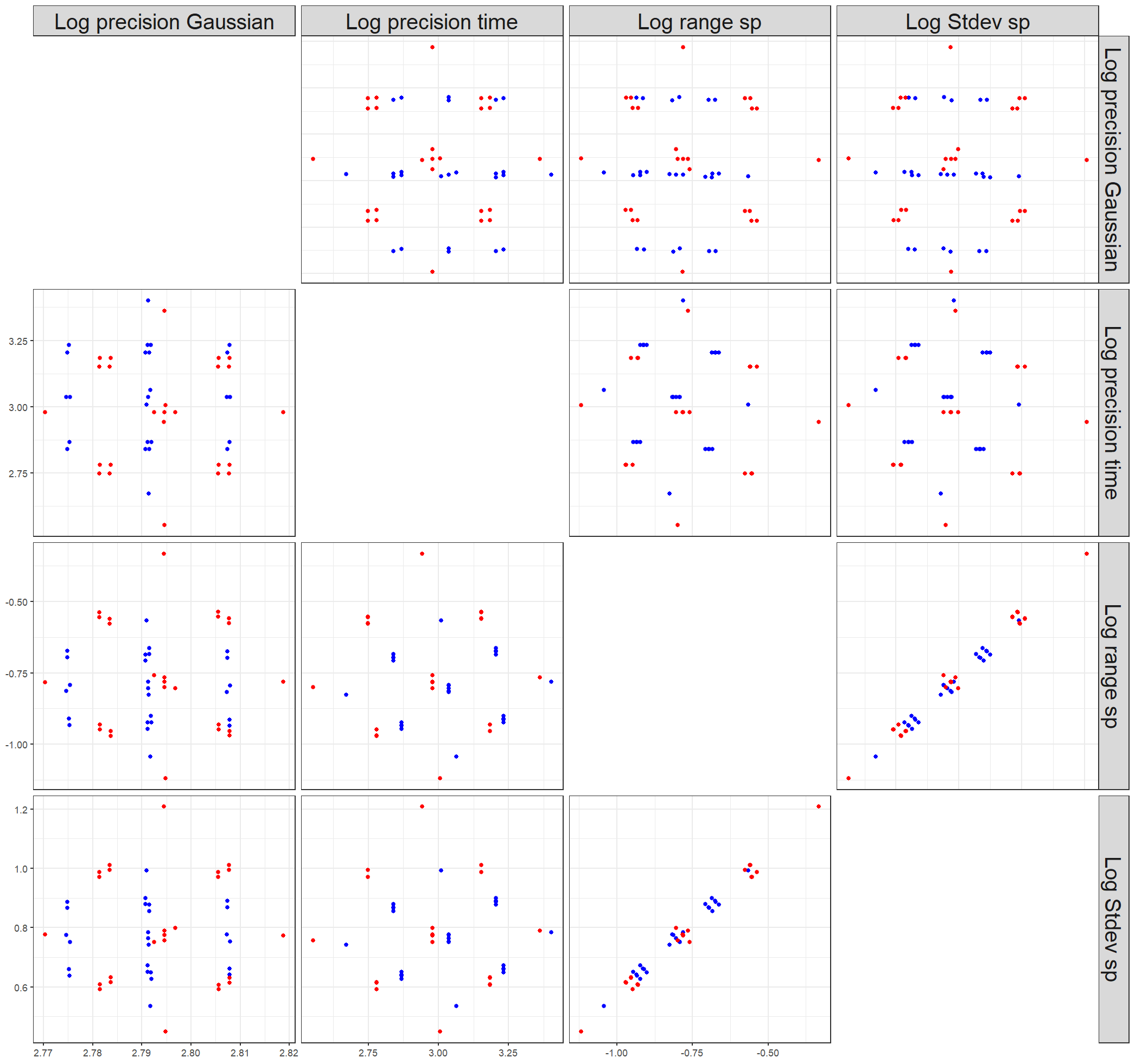}
    \caption{CCD for the full data inference (red) and from the combination of the distributed inference (blue).}
    \label{fig:CCD_simple_example}
\end{figure}

The inference was carried out using the same model as in the simulation. The distributed and recursive approaches were implemented by dividing the data into two partitions, obtained by partitioning the temporal precision matrix of the prior distribution. The CCD integration scheme derived for both approaches is presented in Fig.~\ref{fig:CCD_simple_example}, where the results are shown to be in close agreement with those obtained from the standard full-data analysis.

\begin{figure}
    \centering
    \includegraphics[width=\linewidth]{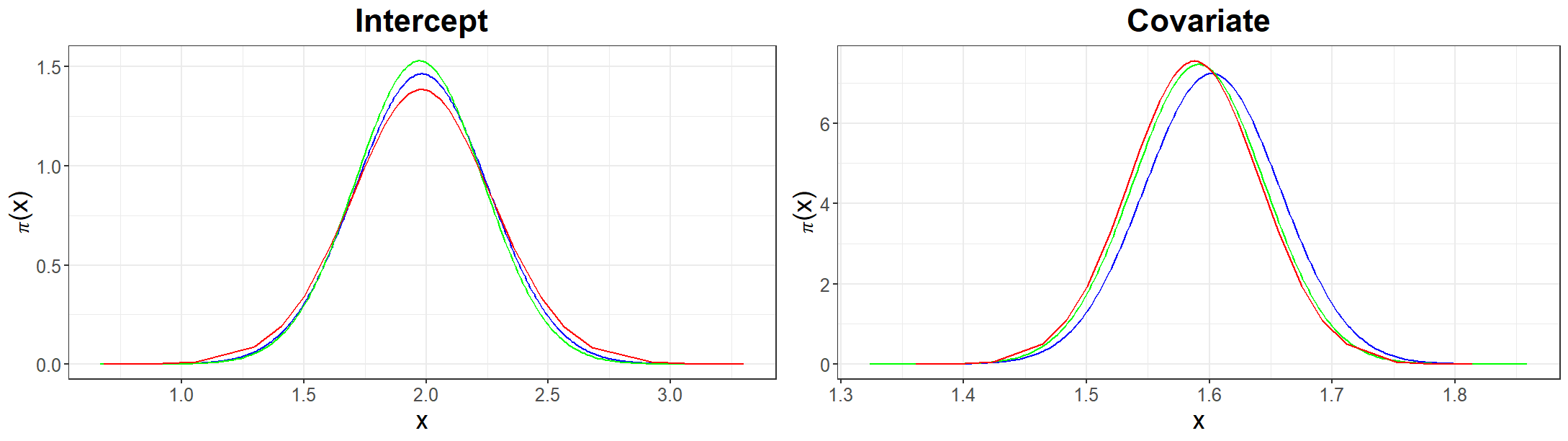}
    \caption{Posterior distributions of the intercept and covariate parameters from the full-data analysis (red), distributed approach (blue), and recursive approach (green).}
    \label{fig:post_fix_simple_example}
\end{figure}

\begin{figure}
    \centering
    \includegraphics[width=\linewidth]{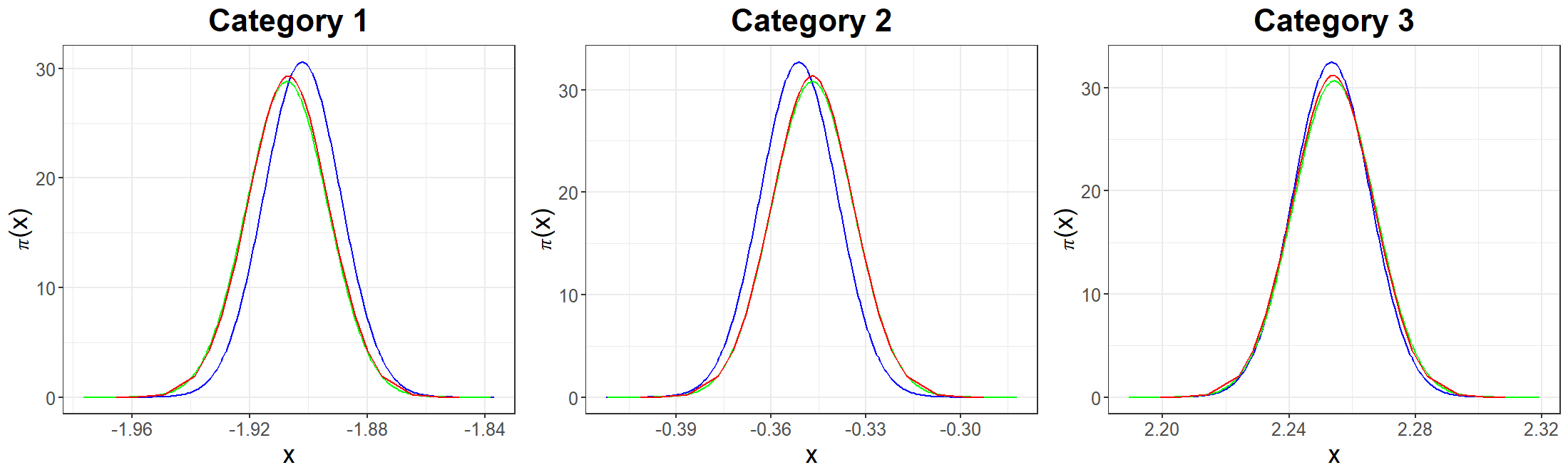}
    \caption{Posterior distributions of the categorical variable parameters from the full-data analysis (red), distributed approach (blue), and recursive approach (green).}
    \label{fig:post_cat_simple_example}
\end{figure}

\begin{figure}
    \centering
    \includegraphics[width=\linewidth]{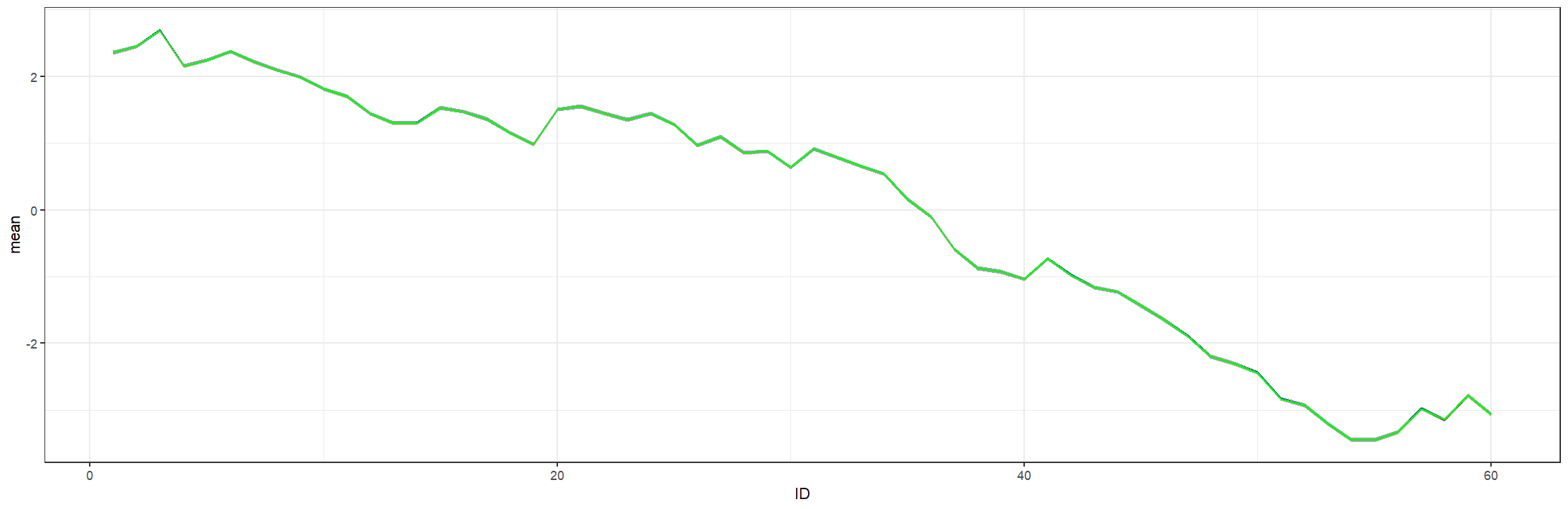}
    \caption{Posterior mean and $95\%$ credible intervals of the temporal effect under the full-data (red), distributed (blue), and recursive (green) analyses.}
    \label{fig:post_temp_simple_example}
\end{figure}

\begin{figure}
    \centering
    \includegraphics[width=\linewidth]{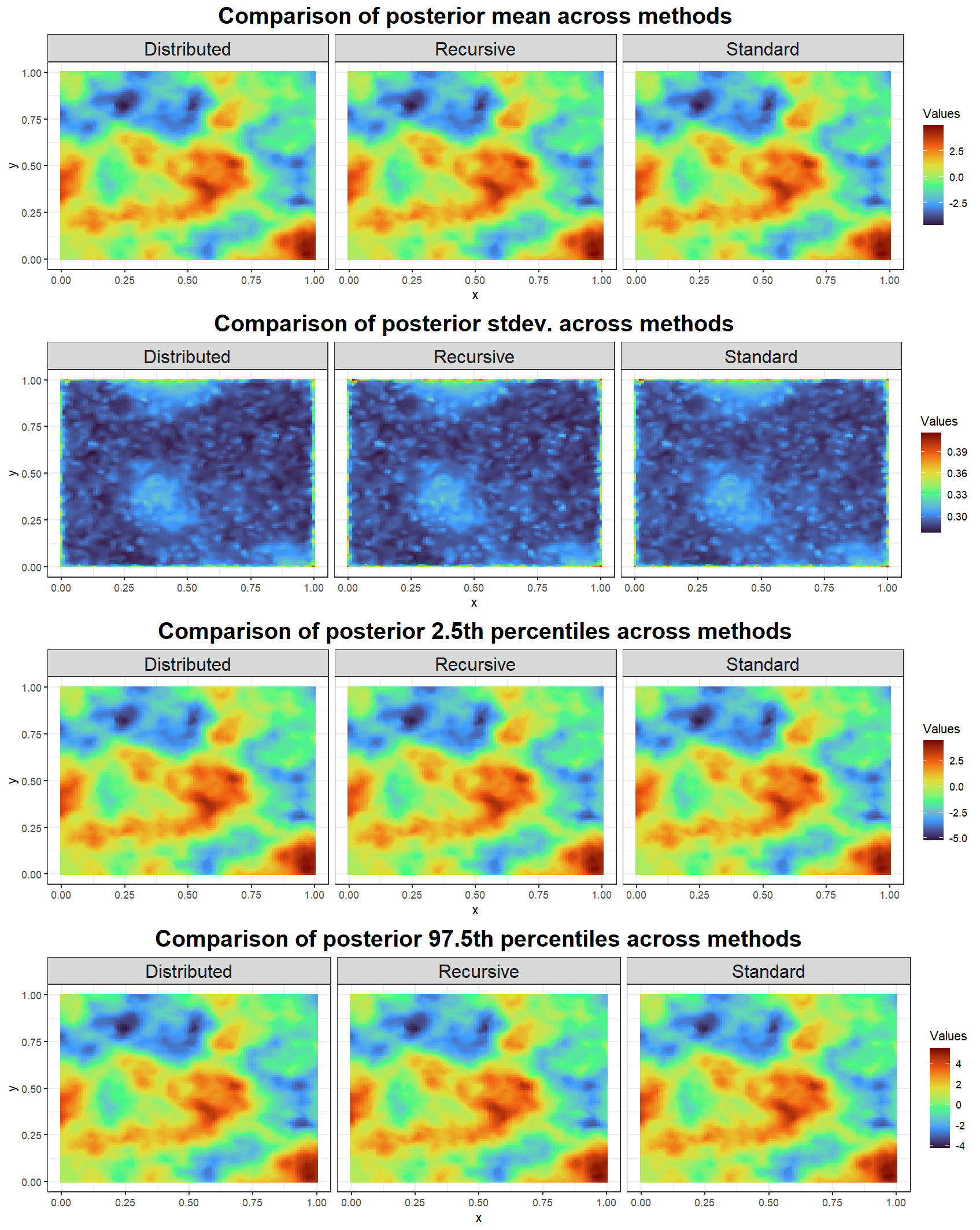}
    \caption{Comparison of the posterior mean, standard deviation, and 2.5th and 97.5th percentiles of the spatial effect across distributed, recursive, and standard full-data analyses.}
    \label{fig:post_sp_simple_example}
\end{figure}

The marginal distributions for the fixed effect and the covariate are displayed in Fig.\ref{fig:post_fix_simple_example}, while Fig.\ref{fig:post_cat_simple_example} reports the corresponding results for the levels of the categorical variable. The temporal trend, together with its $95\%$ credible intervals, is presented in Fig.\ref{fig:post_temp_simple_example}. In turn, Fig.\ref{fig:post_sp_simple_example} illustrates the posterior summaries of the spatial effect, including the posterior mean, standard deviation, and the $0.025$ and $0.975$ quantiles.

\begin{figure}
    \centering
    \includegraphics[width=\linewidth]{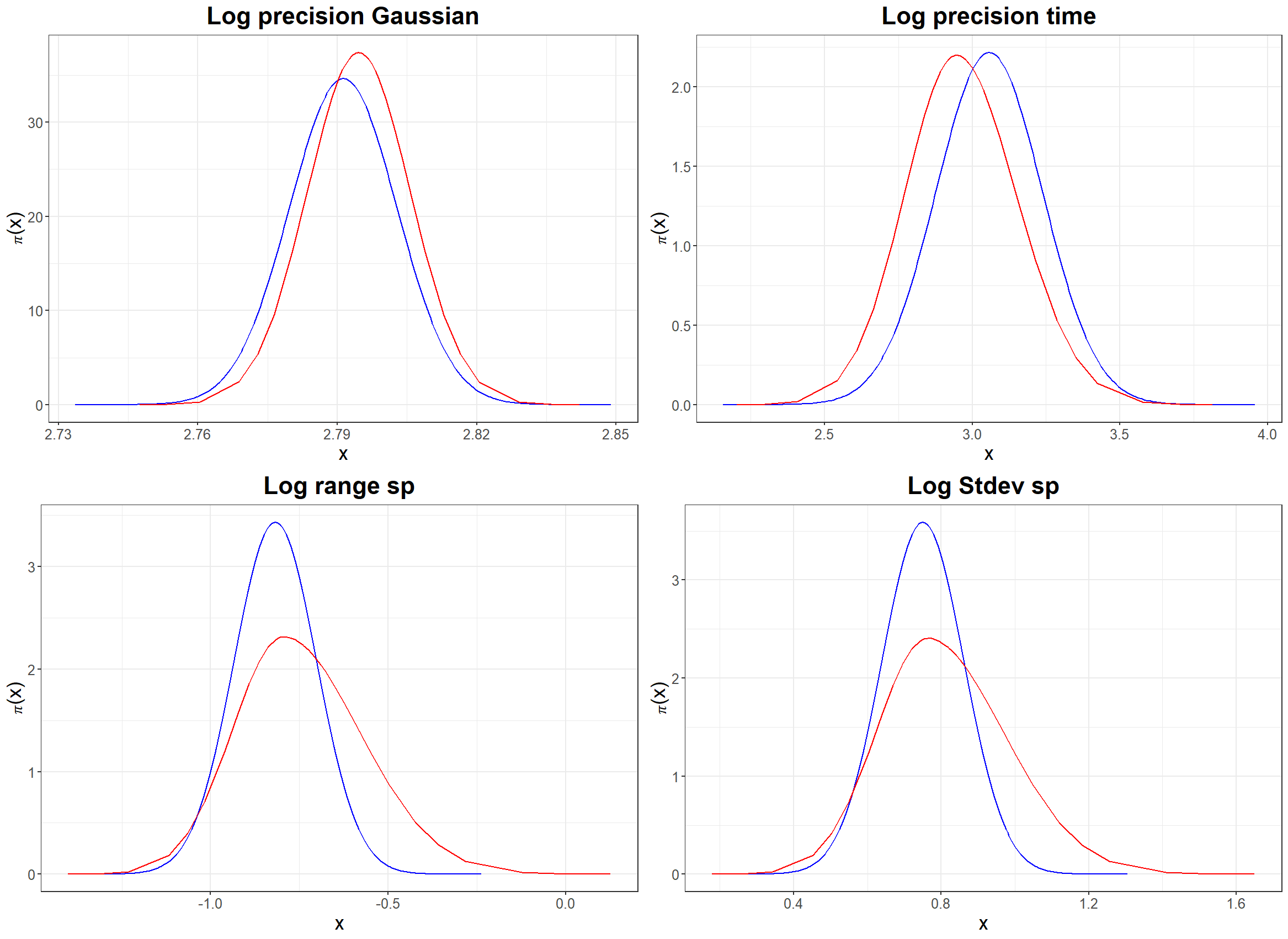}
    \caption{Posterior distributions of the hyperparameters on the internal scale, obtained from the full-data analysis (red) and from the distributed and recursive approaches (blue).}
    \label{fig:post_hyper_simple_example}
\end{figure}

Finally, the marginal posteriors of the hyperparameters for the distributed and recursive approaches were computed using the Gaussian approximation at the mode. These results are compared in Fig.~\ref{fig:post_hyper_simple_example} with those obtained from the standard full-data analysis, where the marginals were calculated using the \textit{numerical integration free algorithm}. 

\subsection{Top-down distributed for a complex spatio-temporal model}

In this example, we analyze a real dataset of PM$_{2.5}$ concentrations across the conterminous United States. Fine particulate matter (PM$_{2.5}$) is one of the six criteria pollutants regulated by the U.S. Environmental Protection Agency (EPA). Data on PM${_2.5}$ mass concentrations and chemical speciation were obtained from the EPA's Air Quality System (AQS) database (\url{https://aqs.epa.gov/aqsweb/airdata/download_files.html}), which compiles measurements from both the CSN and IMPROVE networks. For the spatio-temporal analysis, the dataset was cleaned by treating near-zero negative values as negligible ($10^{-4}$) and removing extreme outliers, both negative and positive. After this preprocessing, the dataset contained $873,762$ annual average observations of PM${2.5}$, expressed in micrograms per cubic meter, collected between 1997 and 2024.

\begin{sidewaysfigure}
    \centering
    \includegraphics[width = \linewidth]{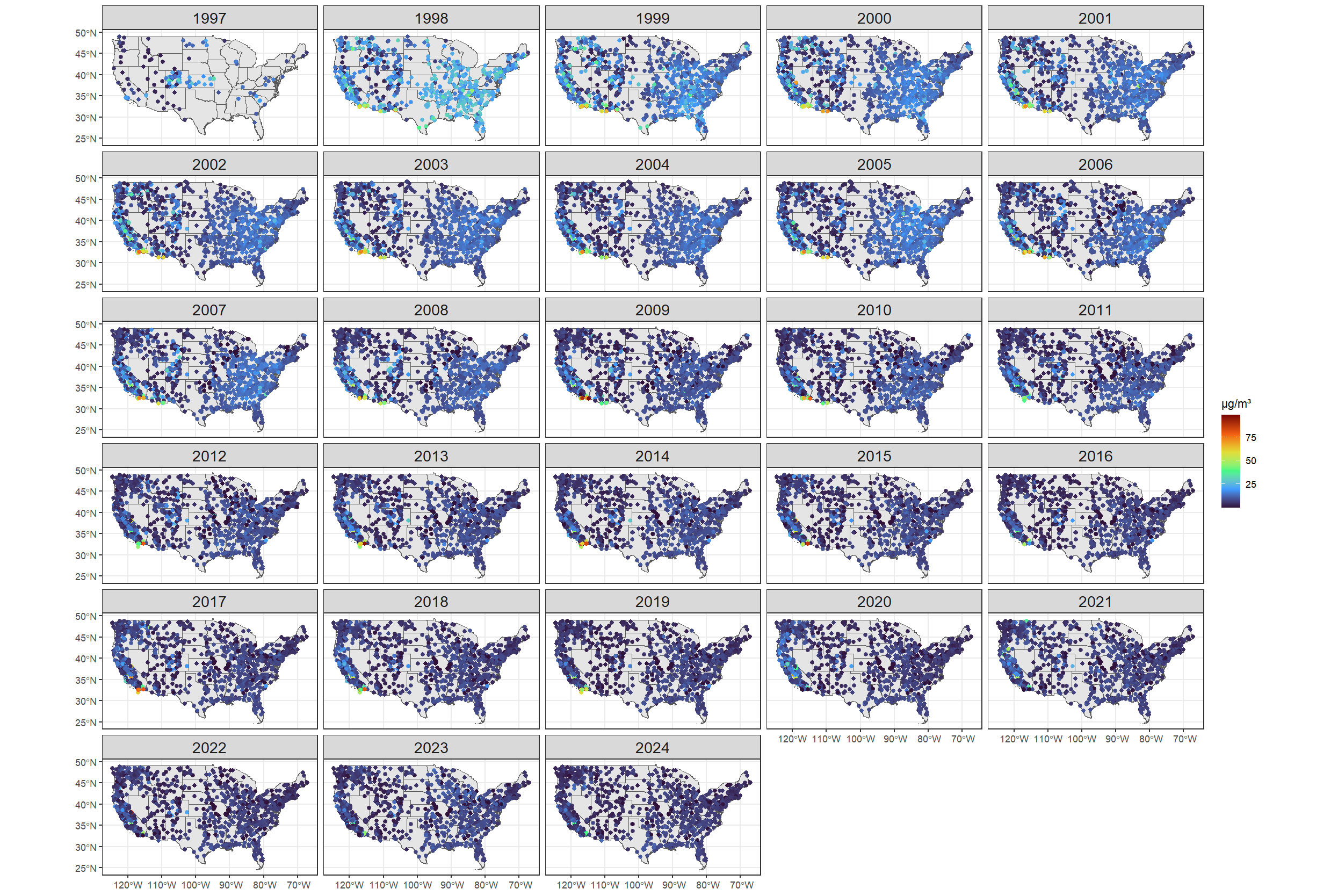}
    \caption{PM$_{2.5}$ data from monitoring stations across the contiguous United States, measured between 1997 and 2024. Values are reported in micrograms per cubic meter $[\mu g/m^3]$.}
    \label{fig:PM25_data_yearly}
\end{sidewaysfigure}

In Fig.~\ref{fig:PM25_data_yearly}, the complete dataset after preprocessing is displayed across the study period. The figure clearly shows the presence of spatial patterns in PM$_{2.5}$ concentrations as well as a temporal trend indicating a decrease in pollutant levels over time. To analyze these data, we employed a model with a Gamma likelihood, as the observations are strictly positive. The latent field consists of a global intercept, a purely temporal component (to capture the clear global temporal trend in the data), and a spatio-temporal interaction component that accounts for variations in the spatial pattern over time:
\begin{equation}
\begin{array}{rcl}
    y_i & \sim & \text{Gamma}(\mu_i, \phi),\\
    \mu_i & = & \beta_0 + u_{ti} + u_{sti},
\end{array}
\end{equation}
where $\beta_0$ is the global intercept, $u_{ti}$ is the purely temporal component modeled as a first-order random walk, and $u_{sti}$ is the spatio-temporal interaction component. The latter is defined as
$$
\mathbf{u}_{st} = \text{GMRF}(\mathbf{0}, \mathbf{Q}_{st}),
$$ 
with precision matrix $\mathbf{Q}_{st} = \mathbf{Q}_t \otimes \mathbf{Q}_s$. That is, a separable spatio-temporal interaction represented by the Kronecker product of the precision matrix of the temporal structure (here based on a first-order autoregressive process) and the precision matrix of the spatial structure (defined using the SPDE-FEM approach). Thus, 
$$
\mathbf{Q}_{st} = \mathbf{Q}_{st}(\sigma_{st}, \rho_{st}, \phi_{st}),
$$
where this separable spatio-temporal component is characterized by three hyperparameters: the marginal standard deviation $\sigma_{st}$, the spatial correlation range $\rho_{st}$, and the temporal autocorrelation parameter $\phi_{st}$. This spatio-temporal interaction component, based on a mesh with $1,211$ spatial nodes and $28$ temporal nodes, results in a precision matrix $\mathbf{Q}_{st}$ of size $33,908 \times 33,908$, which can be demanding in terms of memory usage. 

In this case, the partitioning of both the data and the latent field was also carried out using the structure of the temporal precision matrix $\mathbf{Q}_t$ through Algorithm~\ref{alg:partitioning_bandwidth_reduction}. This approach reduces the dimension of the spatio-temporal interaction component to $1211 \times (14+1)$, where extending to the immediate neighbors of the temporal component is relatively straightforward to handle in the subsequent computations.

\begin{figure}
    \centering
    \includegraphics[width=\linewidth]{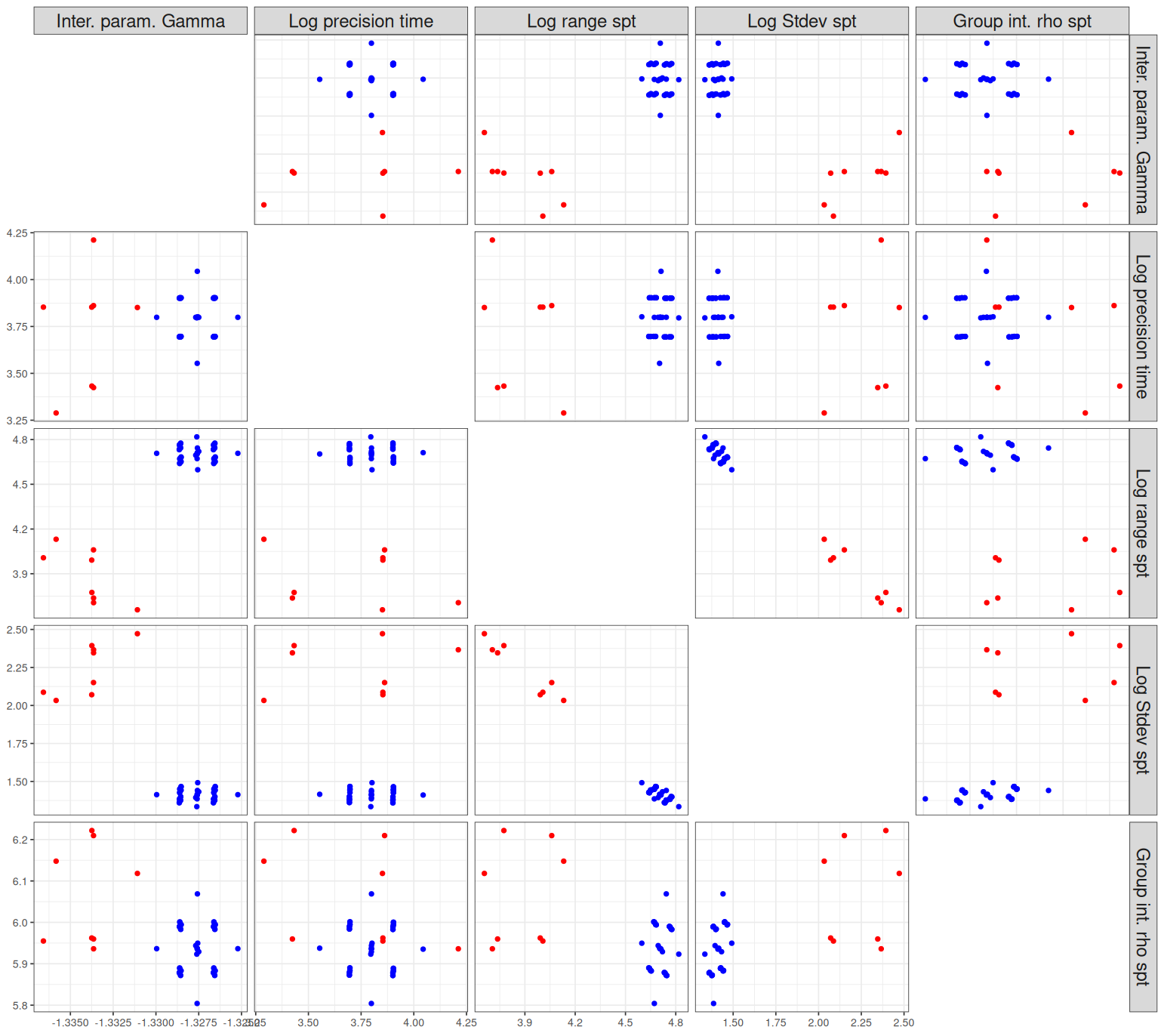}
    \caption{CCD plot of the integration scheme in the hyperparameter space for full data inference (red) and combined distributed inference (blue). Hyperparameters are shown on the internal scale and include: the internal precision parameter of the Gamma likelihood, the log-precision of the temporal component, the log-precision and log-range of the spatio-temporal component, and the internal parameter for temporal autocorrelation in the spatio-temporal interaction.}
    \label{fig:CCD_complex_example}
\end{figure}

The results of the distributed computation and the standard full-data analysis for the CCD integration scheme are presented in Fig.~\ref{fig:CCD_complex_example}. Substantial differences can be observed, which lead to clear discrepancies in the marginal distributions of the hyperparameters. However, these differences do not propagate to the marginals of the latent field, which remain largely consistent across methods. The divergence can arise partly from heterogeneity in the hyperparameters ---for instance, the spatial range may vary across years--- and is further amplified by the non-Gaussian nature of both the hyperparameter posterior and the likelihood. Together, these factors accentuate the gap between the posterior obtained with the standard inferential method and that of the distributed approach.

\begin{figure}
    \centering

    \begin{subfigure}[b]{\textwidth}
        \centering
        \includegraphics[width = 0.9\linewidth]{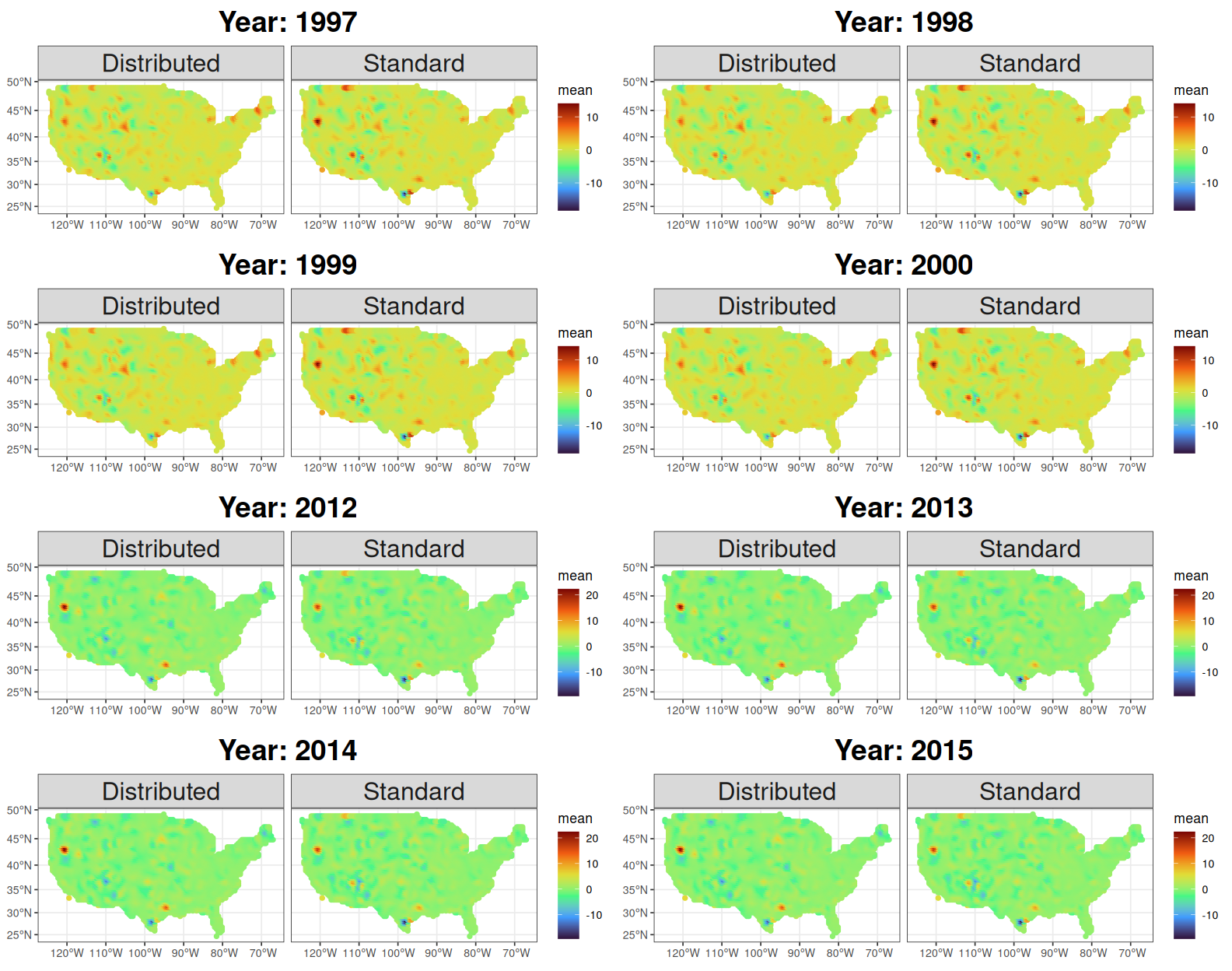}
        \caption{Estimated posterior mean of the spatio-temporal effect.}
    \end{subfigure}
    
    \begin{subfigure}[b]{\textwidth}
        \centering
        \includegraphics[width = 0.9\linewidth]{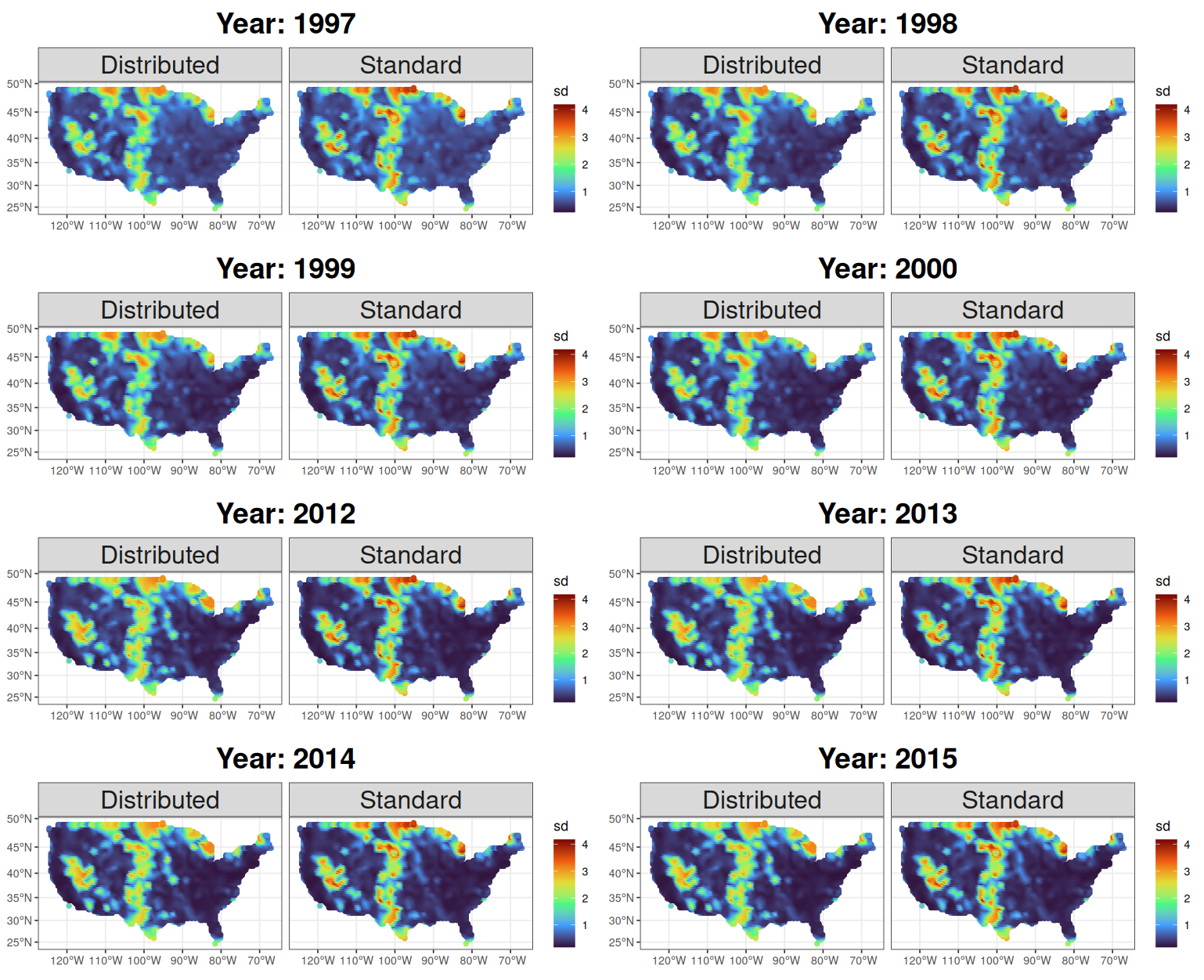}
        \caption{Estimated posterior standard deviation of the spatio-temporal effect.}
    \end{subfigure}
    
    \caption{Comparison of posterior mean and standard deviation of the spatial effect for selected years between distributed and standard full data analyses.}
    \label{fig:spatiotemporal_eff_complex_example}
\end{figure}

\begin{figure}
    \centering
    \includegraphics[width=\linewidth]{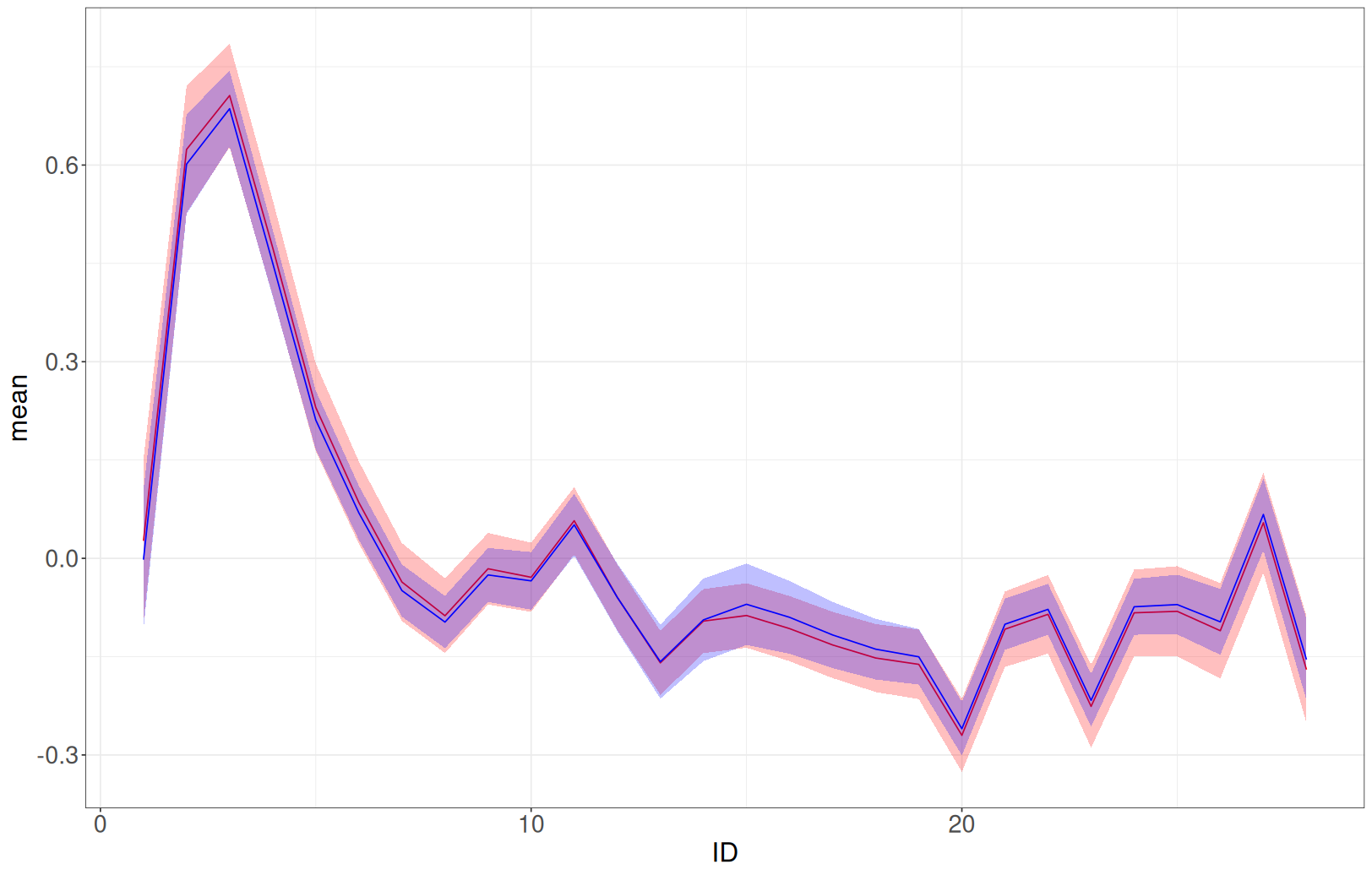}
    \caption{Posterior mean and $95\%$ credible intervals of the temporal effect under the full-data (red) and distributed (blue) analyses.}
    \label{fig:temporal_trend_complex_example}
\end{figure}

\begin{figure}
    \centering
    \includegraphics[width=\linewidth]{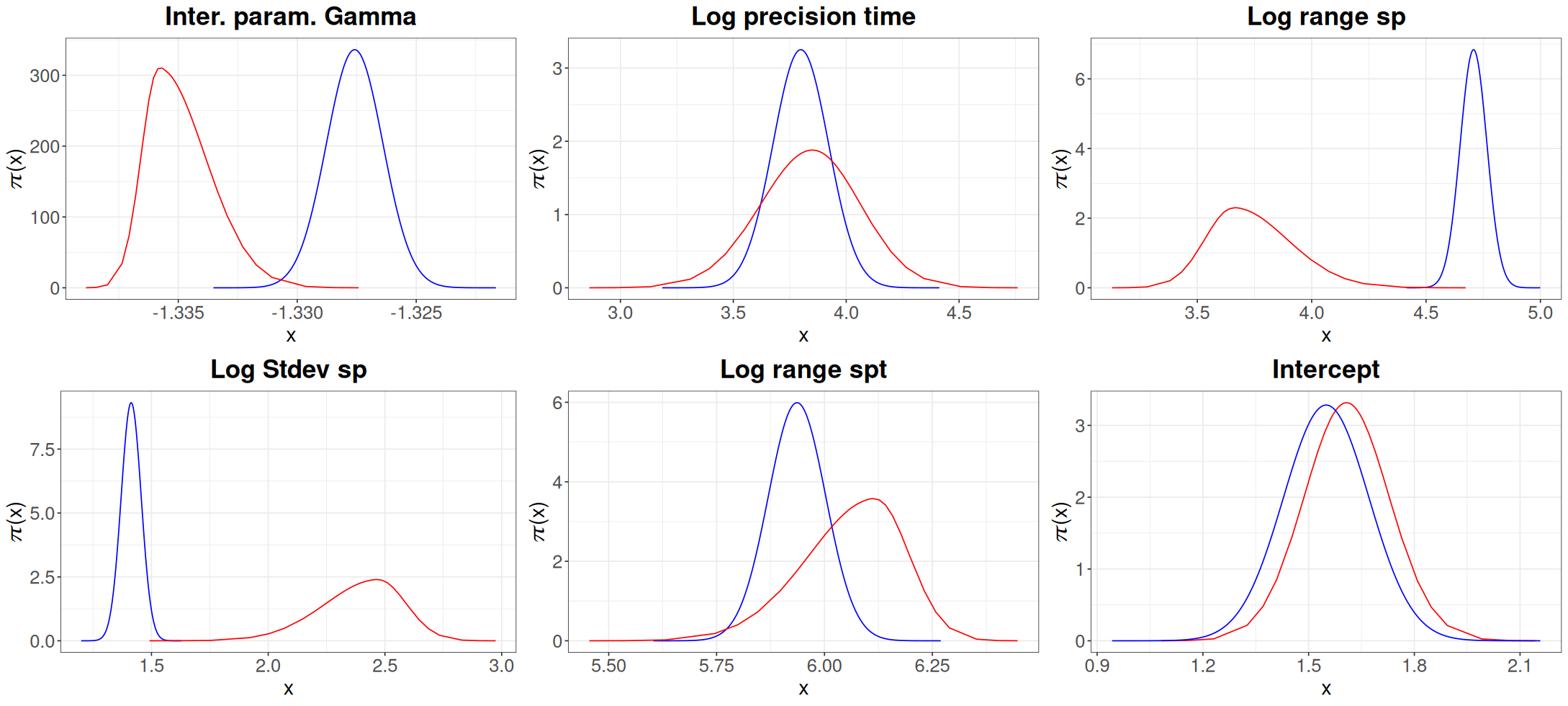}
    \caption{Posterior distributions of the hyperparameters on the internal scale and the posterior of the intercept, obtained from the standard full-data analysis (red) and the distributed method (blue).}
    \label{fig:hyperparam_complex_example}
\end{figure}

Figure~\ref{fig:spatiotemporal_eff_complex_example} displays the mean and standard deviation of the spatio-temporal effect for a subset of years, comparing the results obtained with the distributed method and the standard full-data analysis. The figure shows that both methods yield very similar outcomes. Figure~\ref{fig:temporal_trend_complex_example} presents the mean and the $95\%$ credible interval of the temporal effect for both methods, again indicating close agreement between the two approaches. Finally, Figure~\ref{fig:hyperparam_complex_example} shows the marginal distributions of the hyperparameters on their internal scale, as well as the intercept. While the hyperparameters exhibit noticeable differences in some cases, as previously discussed, the marginal distribution of the intercept remains quite similar across both methods.

\section{Conclusions}

This work has presented a methodology for performing distributed and recursive inference by leveraging the INLA framework. In particular, we have provided a detailed exposition of the different procedures that can be followed to conduct inference under these two paradigms. Special attention has been given to the strategies available for partitioning both the data and the latent field, as well as their implementation within distributed and recursive inferential procedures. The implementation in the \texttt{R-INLA} software has been described to facilitate the application of these methods without the need to program and design every step from scratch, thus making full use of the \texttt{R-INLA} package. Finally, the methodology has been illustrated through its application to both a simulated dataset and a real dataset.

Throughout the development, we have highlighted the advantages of the proposed methods, particularly their ability to partition the latent field and the data jointly, thereby reducing computational costs in terms of both memory and processing time. At the same time, we have discussed the limitations of the approach, which become apparent in the second example when identifying support points and marginal posterior distributions. These limitations stem from the geometry of the joint marginal posterior distribution of the hyperparameters. In cases where hyperparameters could themselves varying along the analyzed data, assuming a single posterior distribution for the entire dataset may introduce divergences. This issue is further compounded by the non-Gaussian nature of the posterior, which amplifies discrepancies when information is transferred through Gaussian approximations around the modal configurations.

Nevertheless, the methodology proves capable of recovering the structure of the latent field, even when it has been partitioned. This constitutes one of its main advantages, as it enables a more flexible combination of information from diverse sources, even when these sources do not share the same latent structure, when certain components of the latent field are not common, or when the hyperparameters differ. The extension of the latent field, together with the modal information of the hyperparameters encoded through a Gaussian approximation, allows the mean vector and precision matrix to be expanded as described in various sections of this article. In other words, these methodologies open the door to addressing problems such as meta-analysis, inference under data-privacy constraints, the integration of heterogeneous information sources (including expert elicitation), and, importantly, big data contexts requiring scalable Bayesian inference.

\clearpage

\vfill
\clearpage

\clearpage





\bibliographystyle{ba}
\bibliography{bibliography}


\end{document}